\documentclass[aps,pre,twocolumn,showpacs,amsfonts]{revtex4}

\usepackage{amsmath, amsthm, amssymb}
\usepackage{amsfonts}
\usepackage{epsfig}
\usepackage{graphicx}
\usepackage{dcolumn}
\usepackage{bm}

\newcommand{\abs}[1]{\left| #1 \right|} % for absolute value
\newcommand{\avg}[1]{\left< #1 \right>} % for average
\renewcommand{\d}[2]{\frac{d #1}{d #2}} % for derivatives
\newcommand{\dd}[2]{\frac{d^2 #1}{d #2^2}} % for double derivatives
\newcommand{\pd}[2]{\frac{\partial #1}{\partial #2}}
\begin{document}
\title{Navier-Stokes hydrodynamics of thermal collapse in a freely cooling granular gas}

\author{Itamar Kolvin}%$^1$}
\author{Eli Livne}%$^1$}
\author{Baruch Meerson}%$^1$}
\affiliation{%$^1$
Racah Institute of Physics, Hebrew University of Jerusalem,
Jerusalem 91904, Israel}
\date{\today }

\begin{abstract}
We employ Navier-Stokes granular hydrodynamics to investigate the long-time behavior of clustering instability in a freely cooling dilute granular gas in two dimensions.
We find that, in circular containers, the homogeneous cooling state (HCS) of the gas loses its stability
via a sub-critical pitchfork bifurcation. There are no time-independent solutions for the gas density in the supercritical region, and we present analytical and numerical evidence that the gas develops thermal collapse unarrested by heat diffusion.
To get more insight, we switch to a simpler geometry of a narrow-sector-shaped container. Here the HCS loses its stability via a transcritical bifurcation. For some initial conditions
a time-independent inhomogeneous density profile sets in, qualitatively similar to that previously found in a narrow-channel geometry. For other initial conditions, however, the dilute gas develops thermal collapse unarrested by heat diffusion. We determine the dynamic scalings of the flow close to collapse analytically and verify them in hydrodynamic simulations. The results of this work imply that, in dimension higher than one, Navier-Stokes hydrodynamics of a dilute granular gas is prone to finite-time density blowups. This provides a natural explanation to the formation of densely packed clusters of particles in a variety of initially dilute granular flows.
\end{abstract}


\maketitle

\section{Introduction}

Finite-time singularities are ubiquitous in hydrodynamic flows, and whenever they appear they attract much interest \cite{Trubnikov,Eggers_review,Kadanoff}. Wave breaking in ideal gas flows \cite{LL}, pinching of droplets \cite{Eggers} and creation of black holes \cite{Penrose}
are examples of finite-time singularities which dramatically change the dynamics in the system when they appear.
Singularities often signal breakdown of theory. In this case either the theory is regularized by introducing additional macroscopic mechanisms, or the continuum description is abandoned in favor of a discrete microscopic theory. A long-standing problem of hydrodynamics is whether a finite-time vorticity singularity develops in an initially smooth three-dimensional incompressible flow. Despite considerable effort, both the inviscid Euler case \cite{Gibbon}, and the viscous, Navier-Stokes case \cite{Doering} remain unresolved.

The present paper deals with a somewhat simpler problem of finite-time density singularities in dilute granular gases. In contrast to ordinary molecular gases, granular gases ``cool" spontaneously because of the inelastic collisions between the grains. Strikingly, even a slight inelasticity of the collisions may cause the granular gas to form dense clusters of particles, even when starting from a dilute and macroscopically homogeneous initial state. This fascinating example of structure formation has been investigated by means of molecular dynamics simulations \cite{Hopkins,Goldhirsch,McNamara2,Grossman,Deltour,Ernst,Puglisi1,Luding,Argentina1,van Noije,Ben-Naim2,MPSS,Saluena,MP,Volfson,PuglisiA} and hydrodynamic simulations \cite{LMS1,LMS2,Bromberg,MPSS,Saluena,Volfson,ELM,Fouxon1,Fouxon2,PuglisiA}. In addition, as many as five different hydrodynamical scenarios have been suggested for the clustering instability; see Ref. \cite{PuglisiA} for a concise critical review.  The clustering instability in granular gases has an interesting analog in the form of  condensation instability in optically thin gases and plasmas that cool by their own radiation \cite{MeersonRMP}.

The Navier-Stokes granular hydrodynamics is a natural starting point for a physicist who wants to understand and predict collective physical phenomena in granular gases \cite{BP,Goldhirsch2}. Granular hydrodynamics provides powerful insights into the physics of granular flow, not in a small part because its predictions are occasionally successful beyond its (quite restrictive) formal limits of validity \cite{Goldhirsch2}.
In the present work we will employ the Navier-Stokes granular hydrodynamics in an attempt to better understand the physics of clustering, including its ultimate stage of thermal collapse, in a freely-cooling dilute granular gas in two dimensions (2D). We will assume throughout the paper that the particle collisions are nearly
elastic,  the local gas density (that we denote by $\rho$) is
much smaller than the close-packing density, and the
Knudsen number of the granular gas is very small:
\begin{equation}\label{threeineq}
    1-\alpha^2\ll 1\,,\;\;\;\;\;\rho \sigma^d\ll 1\,,\;\;\;\;
    \mbox{and}\;\;\;\;\;l_{free}/L\ll1\,.
\end{equation}
Here $0\leq \alpha <1$ is the coefficient of normal restitution of binary
collisions, $\sigma$ is the particle diameter, $d\ge 1$ is the dimension of space, $l_{free}$ is the mean free path of the
gas, and $L$ is the characteristic hydrodynamic length scale. Under these assumptions (the second and third of them need to be verified
\textit{a posteriori}, after the hydrodynamic problem in question is solved) the Navier-Stokes hydrodynamics provides a quantitatively
accurate leading-order theory \cite{BP,Goldhirsch2}.

In a sufficiently small container, a freely cooling granular gas becomes homogeneous because of heat diffusion. From then on, the gas temperature decays according to Haff's law $T\sim (t_0+t)^{-2}$, where $t$ is time, and $t_0=const$ \cite{Haff}.
This Homogeneous Cooling State (HCS) is linearly unstable, however, when the container size exceeds a critical value \cite{Goldhirsch,Deltour,Ernst,McNamara1}. The instability manifests itself by the appearance of density inhomogeneities (the clustering instability) and vortical motions (the shear instability), each of them with its own critical system size.

Preceding our work is a series of papers \cite{ELM,Volfson,Fouxon1,Fouxon2,PuglisiA,MFV} which dealt with nonlinear clustering dynamics of a freely-cooling granular gas in a narrow channel.  Fouxon \textit{et al.} \cite{Fouxon1,Fouxon2} investigated the nonlinear clustering analytically and numerically, using \textit{ideal} (Euler) granular hydrodynamics for a dilute gas which neglects heat diffusion and viscosity. They found a family of exact analytic solutions for the clustering flow. These solutions describe the evolution of an initially smooth flow toward thermal collapse: a finite-time density blowup. Close to the blowup time $t_*$, the maximum gas density exhibits a power law behavior $\sim
(t_*-t)^{-2}$. The minimum temperature of the gas vanishes as  $(t_*-t)^2$. The velocity gradient blows up as $\sim   (t_*-t)^{-1}$, whereas
the velocity itself remains continuous and develops a cusp, rather than a shock discontinuity, at the singularity. Molecular dynamics simulations of a freely cooling granular gas in a channel geometry \cite{PuglisiA} showed an excellent agreement with one of the exact solutions until sufficiently close to the attempted collapse, where heat diffusion becomes important. At finite gravity, a one-dimensional (1D) cooling flow also develops thermal collapse, and exhibits the same asymptotic scaling behavior near the singularity, even in the framework of non-ideal, Navier-Stokes granular hydrodynamics \cite{Volfson}.

Despite the blowup of the density and the velocity gradient, the gas pressure  behaves regularly in all examples of thermal collapse considered in Refs. \cite{Volfson,Fouxon1,Fouxon2}. At zero gravity the collapse occurs at almost constant pressure \cite{Fouxon1,Fouxon2}, whereas at finite gravity the pressure is approximately hydrostatic \cite{Volfson}.  The authors of Refs. \cite{Volfson,MFV} took advantage of these salient features to greatly simplify the full 1D hydrodynamic problem, which takes a full account of heat diffusion and viscosity. It turns out that, at zero gravity, the heat diffusion arrests the thermal collapse in 1D \cite{MFV}. Instead of blowing up, the gas density approaches a time-independent Inhomogeneous Cooling Sate in which the heat diffusion balances the clustering instability brought about by the collisional energy loss.

These 1D hydrodynamic results, however, are strikingly different from those observed in molecular dynamics simulations of a freely cooling granular gas in 2D \cite{Goldhirsch,McNamara2,Ernst,Luding},
where the gas density continues to grow until close-packed clusters of of particles are formed.  To address this issue, the nonlinear hydrodynamic theory of clustering must be extended to higher dimensions. The first step in this direction has been recently taken by Fouxon \cite{Fouxon3}.  Assuming a zero pressure gradient and using Lagrangian coordinates he showed that, in the framework of \textit{ideal} hydrodynamics, thermal collapse persists in any dimension. In Lagrangian coordinates the flow singularity turns out to be identical to the one found in 1D \cite{MFV}.  Fouxon \cite{Fouxon3} also briefly considered the \textit{non-ideal}, Navier-Stokes flow. Here he found inhomogeneous solutions with time-independent density profiles, but left open the question of their stability.

\begin{figure}[ht]
\begin{center}
\includegraphics[scale=0.08]{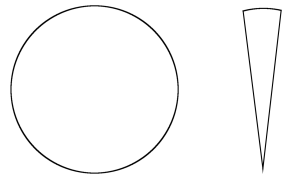}
\end{center}
\caption{Two-dimensional container geometries considered in this work.} \label{Geometries}
\end{figure}

In this work we apply Navier-Stokes hydrodynamics to a freely-cooling granular gas in 2D, in a circular geometry. The final outcome of our work is quite different both from the previous 1D results \cite{MFV}, and from the results of Ref. \cite{Fouxon3}. The hydrodynamic problem is formulated in section \ref{hydro}. In Section \ref{marginal} we perform a marginal stability analysis of the HCS in a circular container and find the critical radii of the container for the clustering modes and shear modes. One of the surprising results, presented in section \ref{disk_section}, shows that the clustering instability appears in this geometry via a sub-critical bifurcation. This suggests that the dilute flow can develop thermal collapse, and we support this conjecture by hydrodynamic simulation (that is, by solving the hydrodynamic equations numerically). To get a more analytical insight we switch, in section \ref{cluster}, to a narrow-sector-shaped container, see Fig. \ref{Geometries}. Here any
flow structure in the azimuthal direction is suppressed by heat diffusion. As a result, the sector geometry renders the simplicity of an effectively 1D (purely radial) time-dependent flow, while retaining some critical 2D features. We find that, in the sector geometry, the bifurcation of the clustering instability is transcritical. Here stable time-independent azimuthally-symmetric inhomogeneous density states exist with a density peak at the circumference of the sector. For these states the density growth is exactly balanced by heat diffusion. But on the same bifurcation diagram there also regions which do not correspond to any time-independent state. Our hydrodynamic simulations in these regions give a strong evidence for thermal collapse unarrested by heat diffusion, this time at the vertex of the sector. We investigate thermal collapse analytically in section \ref{collapse}.  We determine scaling laws, in time and in space, for an \textit{ideal} collapse which we assume
 to develop on the background of an almost constant pressure. Using the scaling laws we find that the non-ideal transport mechanisms -- heat diffusion and viscosity -- are in general unable to arrest the collapse in any dimension higher than one.

\section{Hydrodynamic formulation}
\label{hydro}%

\subsection{Governing equations and boundary conditions}

The granular hydrodynamic equations express local conservation of mass and momentum, and local
energy balance, in the granular gas. Assuming the strong inequalities (\ref{threeineq}), one can use the following set of hydrodynamic equations \cite{BP,Goldhirsch2,Brey}. The continuity equation,
\begin{equation}\label{continuity}
 \pd{\rho}{t} + \nabla\cdot ( \rho \mathbf{v}) = 0,
\end{equation}
and the momentum equation,
\begin{equation}\label{NS}
\rho \left[ \pd{\mathbf{v}}{t} + (\mathbf{v}\cdot\nabla) \mathbf{v} \right] = - \nabla p  + \nu_{0} \nabla \cdot (\sqrt{T} \mathbf{\hat{\Sigma}}). \
\end{equation}
are the usual Navier-Stokes equations for a compressible flow. Here $\rho(\mathbf{r},t)$ is the gas density, $\mathbf{v}(\mathbf{r},t)$ is the gas velocity, and $\nu_{0} \sqrt{T} \mathbf{\hat{\Sigma}}$ is the viscosity tensor. In Cartesian coordinates
\begin{equation}
\hat{\Sigma}_{ki} = \partial_{k} v_{i} +
\partial_{i} v_{k} - \delta_{ik} \mathbf{\nabla}\cdot \mathbf{v},
\end{equation}
see Ref. \cite{LL}. The energy equation reads:
\begin{eqnarray}\label{heat}
\pd{T}{t} + (\mathbf{v}\cdot\nabla) T  &=& -(\gamma - 1) T \nabla\cdot \mathbf{v} + \frac{\kappa_{0}}{\rho} \nabla\cdot ( \sqrt{T}\nabla T)\nonumber \\ &-& \Lambda \rho T^{3/2} + \frac{\nu_{0} (\gamma - 1) \sqrt{T}}{2\rho}\,\mathbf{\hat{\Sigma}}^2,
\end{eqnarray}
where $\gamma$ is the adiabatic index of the gas ($\gamma=2$ and $5/3$ for $d=2$
and $d=3$, respectively), and the ideal-gas equation of state $p = \rho T$ is used. Equations~(\ref{continuity})-(\ref{heat}) differ from the hydrodynamic equations for a dilute gas of \textit{elastically} colliding
spheres only by the presence of the inelastic energy loss term $-\Lambda \rho T^{3/2}$
which is proportional to the average energy loss per collision, $\sim (1-\alpha^2)T$,
and to the collision rate, $\sim \rho T^{1/2}$.
In the viscous heating term of Eq.~(\ref{heat}), $\mathbf{\hat{\Sigma}}^2$ is a scalar. In Cartesian coordinates it is the sum of the squared components $(\hat{\Sigma}_{ik})^2$.

The transport coefficients in Eqs.~(\ref{NS}) and (\ref{heat}) are known from kinetic theory:  $\Lambda=2 \pi^{(d-1)/2} (1-\alpha^2) \sigma^{d-1}/[d\,
\Gamma(d/2)]$ (see \textit{e.g.} \cite{Brey}), where $\Gamma(\dots)$ is the
gamma-function, and $d\ge2$ is the dimension of space, so that $d=2$ corresponds
to disks, and $d=3$ to hard spheres. Furthermore,
$\nu_0=(2\sigma\sqrt{\pi})^{-1}$ and $\kappa_0=4\nu_0$ in 2D, and
$\nu_0=5(3\sigma^2 \sqrt{\pi})^{-1}$ and $\kappa_0=15\nu_0/8$ in 3D \cite{BP}.
Throughout this paper we assume that the particles have a unit mass.

It is often convenient to rewrite the energy equation as an evolution equation for the pressure:
\begin{eqnarray}\label{heatp}
\pd{p}{t} + (\mathbf{v}\cdot\nabla) p  &=& -\gamma p \nabla\cdot \mathbf{v} + \kappa_{0} \nabla\cdot \left[\sqrt{\frac{p}{\rho}}\,\nabla \left( \frac{p}{\rho}\right)\right]\nonumber \\ &-& \Lambda \rho^{1/2} p^{3/2} + \frac{\nu_{0} (\gamma - 1) \sqrt{p}}{2 \sqrt{\rho}}\,\mathbf{\hat{\Sigma}}^2.
\end{eqnarray}
In a circular geometry that we will be dealing with throughout this paper, the natural coordinates are the polar coordinates $r, \theta$. The tensor $\mathbf{\hat{\Sigma}}$ is given in these coordinates by
\begin{eqnarray}
\Sigma_{rr}=\pd{v_r}{r}-\frac{v_r}{r}-\frac{1}{r}\pd{v_\theta}{\theta}\,,
\Sigma_{r\theta}=\pd{v_\theta}{r}+\frac{1}{r}\pd{v_r}{\theta}-\frac{v_\theta}{r}\,.
\end{eqnarray}
As the viscosity tensor is symmetric and traceless \cite{LL}, the other components can be readily found: $\Sigma_{\theta\theta}=-\Sigma_{rr}, \Sigma_{r\theta}=\Sigma_{\theta r}$.

The hydrodynamic equations need to be supplemented by boundary conditions. Let $R$ be the radius of the container.  We choose the boundary conditions so as not to introduce any sources or sinks of mass, momentum and energy, and also not to break azimuthal symmetry. Therefore, we demand a zero granular heat flux through the circumference $r=R$ which, for $T>0$, leads to
\begin{equation}\label{temp_BC}
\pd{T}{r}(r=R,\theta,t) = 0.
\end{equation}
Furthermore, we demand that particle collisions with the wall $r=R$ be elastic, and that the wall be impenetrable. These lead to slip boundary condition for $v_{\theta}$ and zero boundary condition for $v_r$:
\begin{eqnarray}
\label{slip_BC}
  \Sigma_{r\theta} (R,\theta,t) &=& \pd{v_\theta}{r}(R,\theta,t)-\frac{v_\theta(R,\theta,t)}{R} = 0, \nonumber\\
 && v_r (R,\theta,t)= 0.
\end{eqnarray}
Although somewhat restrictive,  boundary conditions (\ref{temp_BC}) and (\ref{slip_BC}) allow a lot of insight into the problem, as will become clear shortly.  The case of no-slip boundary condition, $v_{\theta}(R,\theta,t)=0$,  is briefly discussed in section \ref{marginal} D.

For the narrow-sector-shaped container one should also prescribe boundary conditions on the straight segments of the sector. However, in the absence of azimuthal variations of the hydrodynamic fields, and for $v_\theta \equiv 0$, there is no need of prescribing these boundary conditions explicitly.

\subsection{Global conservation laws}

Equations~(\ref{continuity})-(\ref{heat}), together with the ``conservative" boundary conditions (\ref{temp_BC})-(\ref{slip_BC}), give rise to two conservation laws and a balance relation:
\begin{eqnarray}
\label{mcon}\avg{\rho}  &=& const. \\
\label{Lcon} \avg{\rho  r v_{\theta}} &=& const. \\
\label{Econ} \frac{d}{dt}\avg{ \frac{1}{2} \rho \mathbf{v}^2 + \rho T} &=& - \Lambda\avg{ \rho^2 T^{3/2}},
\end{eqnarray}
where $\avg{\dots}$ denotes averaging over the area of the circular container:
\begin{equation}
    \left\langle \dots\right\rangle = \frac{1}{\pi R^2} \int_D{\dots \,d^2r}\,.
\end{equation}
Equation (\ref{mcon}) describes conservation of the total mass of the gas.
Equation (\ref{Lcon}) describes conservation of the angular momentum of the gas around the center of the container.
This conservation law is special for the circular geometry; it is expected, as neither of the forces in the right hand side of Eq. (\ref{NS}) creates torque. Indeed, a pressure gradient cannot create torque, whereas a viscous stress causes the momentum only to diffuse but not to dissipate.
In addition, the wall $r=R$ exerts, according to the boundary conditions (\ref{slip_BC}),  an effective force on the gas only in the radial direction which again does not produce a torque.

Equation~(\ref{Econ}) describes the global energy loss of the gas.
As the wall is adiabatic, energy can only escape to the internal degrees of freedom of the grains.
Equation~(\ref{Econ}) states that, as long as the gas is not ``frozen" ($T\neq 0$), the
energy will continue to dissipate.
Relations~(\ref{mcon})-(\ref{Econ}) are also valid in the sector geometry. There Eq.~(\ref{Lcon}) is satisfied trivially, as $v_{\theta}=0$.

\section{Marginal stability of the Homogeneous Cooling State}
\label{marginal}

Apart from inertia of the flow, there are two competing mechanisms which ultimately determine the spatial structure of a freely cooling granular gas. These are the inelastic energy loss which drives the clustering- and shear-mode instabilities \cite{Goldhirsch,McNamara1}, and the heat diffusion and viscosity which tend to erase inhomogeneities in the gas. The rest of parameters being fixed,  the heat diffusion and viscosity prevail in sufficiently small containers; here the granular gas approaches the HCS.  When the container size is increased, larger-scale perturbations become possible, and the HCS eventually loses its stability. In this section we will find the critical radii of the circular container for the clustering modes and the shear mode which can bifurcate from the HCS. When the HCS is no longer stable, it is the further development of these modes which determine the late-time behavior of the gas.

Linear stability analysis of the HCS was performed by McNamara \cite{McNamara1}, Deltour and Barrat \cite{Deltour}, Brey \textit{et al.} \cite{Brey} and other workers. Here we extend the previous analysis, that was done in rectangular geometries, to the circular geometry. As we are only interested in the bifurcation points (where the system's behavior qualitatively changes) and do not need to know the  growth rates of linear perturbations, we will limit ourselves to a marginal stability analysis.

\subsection{The HCS is stable in small systems}
\label{HCS_stable}
The HCS is the simplest exact solution of Eqs. (\ref{continuity})-(\ref{slip_BC}). Here $\rho=\rho_0=const$, and $\mathbf{v}=0$. The energy equation  becomes
\begin{equation}
\label{Haff_diff} dT(t)/dt = - \Lambda \rho_0 T^{3/2}(t)\,,
\end{equation}
and so the gas temperature decays according to Haff's law \cite{Haff}:
\begin{equation}
T_H (t) = \frac{4}{[\Lambda \rho_0(t+t_0)]^2}\,,
\end{equation}
where $t_0=const$.
For sufficiently small circular containers, see below,  and if the total angular momentum of the gas is zero, the HCS is a global attractor of the flow in this geometry. On a qualitative level, the HCS stability in small containers is evident from the hydrodynamic equations.  If the system size is $R$, the heat-diffusion term in Eq. (\ref{heat}) scales as $R^{-2}$ while the cooling term is $R$-independent. Therefore, as $R$ becomes small, the heat diffusion generates a flow which erases any inhomogeneities.

\subsection{Small-amplitude modes are decoupled}

Although the HCS is not a steady state of the system in the usual sense, it does become a steady state in rescaled variables. The rescaled temperature is $\tau(\mathbf{r},t) =T(\mathbf{r},t)/T_H (t)$; the density remains unchanged. Marginal stability for the gas velocity implies a constant Mach number: the velocity decays neither faster nor slower than the (time-dependent) speed of sound $\sqrt{\gamma T_H(t)}$. Therefore, we introduce the rescaled velocity $\mathbf{u}$ by putting
\begin{equation}\label{u}
\mathbf{v}(\mathbf{r},t) = \sqrt{\gamma T_H(t)} \mathbf{u}(\mathbf{r},t).
\end{equation}
We emphasize that we will only deal with zero-angular-momentum flows. Indeed, as the HCS has a zero angular momentum, no finite-angular-momentum state can bifurcate from it.

Let us linearize the hydrodynamic equations~(\ref{continuity})-(\ref{heat}) around the HCS. We introduce small perturbations to each of the variables,
\begin{eqnarray}
\rho &=& \rho_0 (1+\delta\rho)\nonumber\\
T &=& T_H (t) ( 1 + \delta\tau)\nonumber\\
\mathbf{v} &=& \sqrt{ \gamma T_H(t)} \delta\mathbf{u} ,
\end{eqnarray}
where $\rho_0$ is the average density of the gas. For marginally stable states
\begin{equation}\label{marg}
    \pd{}{t}\delta\rho =\pd{}{t}\delta\tau=\pd{}{t}\delta\mathbf{u}=0.
\end{equation}
As a result, the continuity equation (\ref{continuity}) becomes
\begin{equation}
\label{marginal_continuity}
\mathbf{\nabla}\cdot\delta\mathbf{u} = 0.
\end{equation}
Next, we linearize the energy equation (\ref{heat}):
\begin{equation}
\label{marginal_heat}
2\delta\rho+\delta\tau-l_{\kappa}^2\nabla^2\delta\tau=0,
\end{equation}
where $l_{\kappa}=\sqrt{2\kappa_0/(\Lambda \rho_0^2)}$ is the critical
length for clustering instability in a rectangular geometry  \cite{Goldhirsch,McNamara1,Brey,MFV}. Finally, we linearize the momentum equation (\ref{NS}):
\begin{equation}
\label{marginal_NS}
\delta\mathbf{u}+l_{\nu}^2\nabla^2\delta\mathbf{u} = \frac{2}{\sqrt{\gamma}\Lambda\rho_0}\mathbf{\nabla}(\delta\rho+\delta\tau),
\end{equation}
where $l_{\nu}=\sqrt{2\nu_0/(\Lambda \rho_0^2)}$ is the critical length for the shear instability in a rectangular geometry \cite{Goldhirsch,McNamara1,Brey,Brey2}. By virtue of Eq. (\ref{marginal_continuity}), the linearized flow is incompressible. As a result,  the velocity field is purely solenoidal. Now we see that the left hand side of Eq. (\ref{marginal_NS}) is solenoidal, while the right hand side is potential. By the Helmholtz decomposition theorem for vector fields, see \textit{e.g.} \cite{Helmholtz}, each side of the equation must be equal to zero. In other words, the shear modes and the density modes (the clustering modes) decouple. As a result,
\begin{equation}
\delta\rho+\delta\tau  = const.
\end{equation}
Now, $\int_D \delta \rho \, d^2 r = 0 $ by virtue of mass conservation, and so  $\int_D \delta \tau d^2 r  = 0$ by virtue of Eq. (\ref{marginal_heat}), where D represents the circular container. Therefore, $\delta\rho=-\delta\tau$: at marginality the small density and temperature perturbations are isobaric. For the velocity we can introduce the Stokes' stream function, as the flow is incompressible. Putting $\delta u_x = \partial \psi/\partial y$ and $\delta u_y=-\partial \psi/\partial x$, we reduce the marginal stability equations to two independent Helmholtz equations
\begin{eqnarray}
\delta\rho+l_{\kappa}^2\nabla^2\delta\rho &=& 0\,, \label{marginal_final_rho}
\\
\psi+l_{\nu}^2\nabla^2\psi &=& 0\,,\label{marginal_final_psi}
\end{eqnarray}
for the clustering modes and shear modes, respectively.

\subsection{Clustering modes}
\label{marginal_modes}

\begin{figure}[ht]
\includegraphics[scale=0.5]{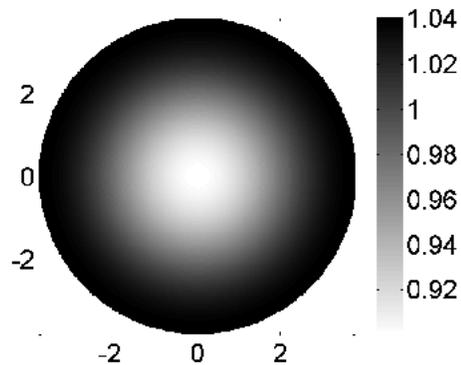}
\caption{The density field $\rho(r,\theta) = 1-0.1\, J_0(r/l_\kappa)$ for the lowest marginally stable clustering mode $m=0$ with $R=R_0=3.83171\dots \, l_\kappa$.}
\label{marginal0}
\end{figure}

\begin{figure}[ht]
\includegraphics[scale=0.5]{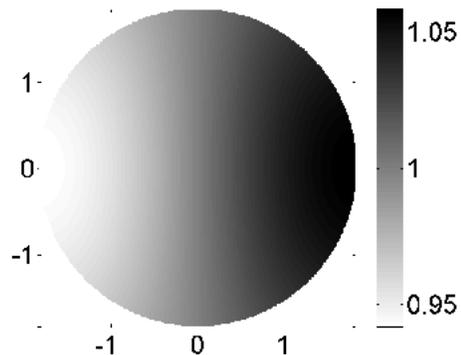}
\caption{The density field $\rho = 1+0.1 \,J_1(r/l_\kappa) \cos (\theta)$  for the lowest marginally stable clustering mode $m=1$. $R=R_1 = 1.84118\dots\, l_\kappa$. }
\label{marginal1}
\end{figure}

The regular independent solutions of Eq. (\ref{marginal_final_rho}) are, up to a constant pre-factor,
\begin{equation}
J_m (r/l_{\kappa}) \left\{\begin{array}{cc}
\cos(m\theta)\cr
\sin(m\theta)
\end{array} \right\},\,\, m =0,1,2, \dots ,
\end{equation}
where $J_m(\dots)$ is the Bessel function of the first kind. To satisfy the boundary condition (\ref{temp_BC}), we demand  $J_m^{\prime}(R/l_{\kappa}) = 0$ and obtain a doubly infinite series of solutions with $m=0,1, \dots$ and quantized values of $R$.
The first two marginally stable modes are the lowest $m=0$ mode with $R = R_0 = 3.83171\dots \, l_{\kappa}$ and the lowest $m=1$ mode with $R=R_1 = 1.84118\dots\, l_\kappa$ (see \textit{e.g.} Ref. \cite{Abramowitz} for the roots of the Bessel functions and their derivatives).
The $m=0$ modes are azimuthally symmetric. The gas density, corresponding to the lowest of them, is $\rho=1+ C_0 J_0 (r/l_\kappa)$, where the sign of the constant $C_0$ determines whether the gas is denser at the center or at the circumference.  An example of this density profile is depicted in Fig. \ref{marginal0}.
For the $m=1$ modes the azimuthal symmetry is broken; the density profile of the lowest of them is $\rho = 1+C_1 J_1(r/l_\kappa) \cos (\theta)$, see Fig. \ref{marginal1}.

As the container radius increases from zero, the first bifurcation occurs at $R=R_1$ for the lowest $m=1$ mode. Therefore, the lowest $m=1$ mode, rather than the lowest $m=0$ mode, is the most unstable clustering mode in this geometry. In the following we will sometimes omit the word ``lowest" in this context.

\subsection{Shear modes}
\begin{figure}[ht]
\includegraphics[scale=0.5]{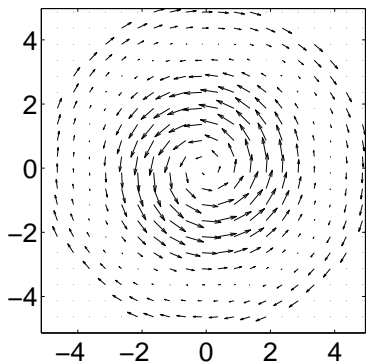}
\caption{The velocity field $\delta \mathbf{u} =J_1(r/l_\nu)\mathbf{\hat{\theta}}$ of the lowest marginally stable shear mode. $R= 5.13562\dots\, l_\nu$. Evident is a zonal flow.}
\label{shearpic}
\end{figure}

Now we proceed to Eq.~(\ref{marginal_final_psi}). The independent regular solutions for the stream function are
\begin{equation}
J_m(r/l_\nu)\left\{\begin{array}{cc}
\cos(m\theta)\cr
\sin(m\theta)
\end{array} \right\},\,\, m =0,1,2, \dots.
\end{equation}
Correspondingly, the velocity components are
\begin{equation}
\begin{array}{cc}
\delta u_r = \frac{1}{r}\pd{\psi}{\theta} = \frac{m}{r} J_m(r/l_\nu) \left\{\begin{array}{cc}
-\sin(m\theta)\cr
\cos(m\theta)
\end{array} \right\},\cr
\delta u_\theta  = - \pd{\psi}{r} = - \frac{1}{l_\nu}J_m'(r/l_\nu) \left\{\begin{array}{cc}
\cos(m\theta)\cr
\sin(m\theta)
\end{array} \right\}.
\end{array}
\end{equation}
The boundary conditions (\ref{slip_BC}) for the velocity become
\begin{equation}
\begin{array}{cc}
J_m(R/l_\nu) = 0 \ \  \mbox{and} \ \ J_m^{\prime}(R/l_\nu) = 0 \ \ \mbox{for} \ \ m=1,2, \dots\,, \cr\cr
J_2(R/l_\nu) = 0 \ \ \mbox{for} \ \ m = 0.
\end{array}
\end{equation}
As the first two conditions, for $m=1,2, \dots$, cannot be satisfied simultaneously, the only possible type of shear modes are the azimuthally symmetric modes $m = 0$, for which the velocity field is
\begin{equation}
\delta u_r = 0 \ \; \mbox{and} \ \; \delta u_\theta = J_1(r/l_\nu).
\end{equation}
The critical radius for the bifurcation of the lowest $m=0$ shear mode is $R_s=5.13562\dots \, l_\nu$. For a comparison with the clustering modes we need to express the critical radius for the $m=0$ shear mode in units of $l_\kappa$. For a gas of hard disks $R_s = 2.56781\dots\, l_\kappa$, whereas for a gas of hard spheres $R_s = 3.75053\dots\, l_\kappa$.
Therefore, one always has $R_s > R_1$, and the clustering mode bifurcates, in the circular geometry, at a smaller container radius than the shear mode. This is in contrast to what happens in rectangular geometries, where a shear mode goes unstable first \cite{McNamara1,Brey,Brey2}.  An interesting feature of the shear modes in a circular container is that they exhibit zonal flows.  This is because the flow must have a zero angular momentum, as imposed by Eqs.~(\ref{Lcon}), (\ref{u}) and (\ref{marg}). One example is shown in Fig. \ref{shearpic} which depicts the velocity field of the lowest $m=0$ mode.

For the no-slip condition at $r=R$ the analysis is very similar. Here the critical radius $R_s$ for the shear mode is determined from equation $J_1(R/l_\nu) = 0$. As a result,
$R_s = 1.91585\dots l_\kappa$ for hard disks, and $R_s = 2.79828\dots l_\kappa$ for hard spheres. Both values exceed the critical radius $R_1 = 1.84118\dots l_\kappa$ for inhomogeneous clustering. Therefore, for the no-slip boundary, the first bifurcation to occur as the system size is increased is again the clustering mode bifurcation. Notably, a zonal flow is absent for the no-slip boundary.

\section{Nonlinear clustering in a circular container}
\label{disk_section}
As we have just found, the clustering mode $m=1$ is the first mode to become unstable as the container radius $R$ exceeds  $R_1 = 1.84118\dots\, l_\kappa$. One possible scenario of the evolution of the unstable $m$=1 mode is the formation of an inhomogeneous density distribution of the gas with no mean flow, $\mathbf{v}=0$. This state can continuously bifurcate from the HCS. In view of Eq.~(\ref{continuity}) this density distribution must be time-independent. In addition, from Eq. (\ref{NS})  the pressure gradient must be zero, $\nabla p = 0$. The density distribution  $\rho(\mathbf{r})$ and the evolution in time of the pressure $p(t)$ are governed by the energy equation (\ref{heatp}) which becomes
\begin{equation}\label{heatsimple}
\dot{p}= -\gamma p \nabla\cdot \mathbf{v} + \kappa_{0} p^{3/2}\nabla\cdot \left[\frac{1}{\sqrt{\rho}}\,\nabla \left( \frac{1}{\rho}\right)\right]-\Lambda \rho^{1/2} p^{3/2},
\end{equation}
with Eq.~(\ref{temp_BC}) as the boundary condition. It is convenient to rescale the variables: the pressure by its characteristic initial value $p_0$, the density by the average density $\rho_0$, the coordinates $x,y$ by the critical length $l_\kappa$ which appears in Eq.~(\ref{marginal_heat}), and the time by the characteristic cooling time
\begin{equation}
t_c = \frac{2}{\Lambda \rho_0^{1/2} p_0^{1/2}}.
\end{equation}
The rescaled problem is the following:
\begin{eqnarray}\label{ICS_eq}
\frac{\dot{p}}{p^{3/2}} = - 2 \rho^{1/2} &+& \nabla\cdot \left (\sqrt{\frac{1}{\rho}}\,\nabla \frac{1}{\rho} \right), \\
\label{ICS_bc} \pd{\rho}{r} \left(r=R\right) &=& 0,
\end{eqnarray}
where we abuse notation by denoting the rescaled quantities by the same symbols as the physical ones. Averaging Eq. (\ref{ICS_eq}) over the (rescaled) area of the container, we obtain a global balance equation for the pressure
\begin{equation}
\label{pressure_law}
\frac{\dot{p}}{p^{3/2}} = -2 \langle\rho^{1/2}\rangle,
\end{equation}
where $\langle \dots\rangle$ denotes averaging over the area of the circular container.
Thus, the time-independent density profiles are determined from the nonlocal problem
\begin{eqnarray}\label{ICS_density}
\frac{1}{3} \nabla^2 \rho^{-3/2} &+& \langle\rho^{1/2}\rangle- \rho^{1/2} = 0,\nonumber\\
\pd{\rho}{r} (r=R) &=& 0.
\end{eqnarray}
This problem is a generalization to 2D of the similar problem for a narrow-channel flow \cite{MFV}. A 3D version of this equation was derived by Fouxon \cite{Fouxon3}. He discussed spherically-symmetric solutions with a single density maximum at $r=R$. In 2D these correspond to the lowest $m=0$ mode, see Fig. \ref{marginal0}. As the $m=1$ mode actually goes unstable first in a circular container, any azimuthally-symmetric solution of Eq.~(\ref{ICS_density}) must be unstable with respect to perturbations breaking this symmetry. We checked that a similar feature persists in 3D: the first mode to become unstable as the radius of a \textit{spherical} container is increased from zero is  \textit{not} spherically symmetric. As a result, in a spherical container, the spherically-symmetric solutions reported in Ref. \cite{Fouxon3} must be unstable with respect to perturbations breaking this symmetry. The azimuthally- and spherically-symmetric solutions (in 2D and 3D, respectively) be
 come, however, relevant in narrow-sector (correspondingly, narrow-cone) shaped containers, where heat diffusion erases any non-radial structure. We will deal with the narrow-sector geometry later on, in section \ref{cluster}. In this section we continue to deal with a circular container. First of all, we will perform a weakly nonlinear analysis of Eq.~(\ref{ICS_density}) to determine the character of bifurcation of the HCS. As we will see, the bifurcation at hand is a pitchfork bifurcation, where the state parameter (a density contrast defined below) grows as a square root of the difference between the container radius and the critical radius. A pitchfork bifurcation can be either super-critical, or sub-critical \cite{Crawford}, thus determining the stability of states below and above the bifurcation. Since this important characteristic is determined by the sign of a numerical coefficient, an accurate calculation is needed to ascertain it.

\subsection{Bifurcation analysis}

Equation~(\ref{ICS_density}) can be simplified by a transformation of the unknown function $\rho$ and of the rescaled radial coordinate $r$:
\begin{equation}\label{Phi_trans}
    \rho= A^2 \Phi^{-2/3},\;\;\;\;\; r=A^{-2} \tilde{r},
\end{equation}
where $A = \langle\rho^{1/2}\rangle$. We obtain a parameter-free nonlinear Poisson equation with a Neumann boundary condition at $\tilde{r} =\tilde{R}\equiv A^2 R$:
\begin{eqnarray}
\label{disk_trans}
\frac{1}{3}\nabla^2\Phi+1-\Phi^{-1/3}&=&0,\nonumber\\
\pd{\Phi}{\tilde{r}}(\tilde{R},\theta) &=& 0.
\end{eqnarray}
Once $\Phi$ is found, $A$ can be determined  from
$A^{-2}=\langle\Phi^{-2/3}\rangle_{\tilde{r}}$,
where the averaging is over the circular container of rescaled radius $\tilde{R}$. The latter relation
follows from the condition $\langle \rho \rangle=1$ for the rescaled density $\rho$.

In the new coordinate $\tilde{r}$,  the critical radius for instability of the mode $m=1$, found in section \ref{marginal_modes}, is $R_1 =  \tilde{R}_1=1.841\dots$. Let us consider a circular container with a radius slightly above or below $R_1$ and seek for a weakly-nonlinear solution of Eq.~(\ref{disk_trans}). At the risk of irritating the reader, we will transform  the radial coordinate one more time, by introducing
$\chi=(\tilde{R}_1/\tilde{R})\tilde{r} $, and also define the parameter $k = \tilde{R}/\tilde{R}_1$ which will serve as the eigenvalue of the nonlinear eigenvalue problem
\begin{eqnarray}
\label{disk_problem_conv}
\frac{1}{3}\nabla^2\Phi+k^2(1-\Phi^{-1/3})&=&0\nonumber\\
\pd{\Phi}{\chi}(\tilde{R}_1,\theta) &=& 0.
\end{eqnarray}
Now we expand both the eigenfunction and the eigenvalue in a series employing a perturbation amplitude $\epsilon$ as a small parameter:
\begin{eqnarray}
\Phi(\chi,\theta) &=& 1 + \epsilon \varphi_1(\chi,\theta) + \epsilon^2 \varphi_2(\chi,\theta)\nonumber\\
&+&\epsilon^3\varphi_3(\chi,\theta)+\dots\nonumber\\
k &=& 1 + \epsilon k_1+\epsilon^2 k_2+\dots,
\end{eqnarray}
where, according to our marginal stability analysis, $\varphi_1 = J_1(\chi) \cos(\theta)$. In the second order in $\epsilon$ we obtain
\begin{eqnarray}
\label{disk_problem_2nd}
\nabla^2\varphi_2+\varphi_2&=&\frac{2}{3}\varphi_1^2-2k_1\varphi_1\nonumber\\
\pd{\varphi_2}{\chi}(\tilde{R}_1,\theta) &=& 0.
\end{eqnarray}
The solvability condition of the inhomogeneous linear equation is the orthogonality of its right hand side to the function $\varphi_1$. This yields $k_1 =0$. Solving the resulting equation by a Fourier decomposition, we find
\begin{equation}
\varphi_2 = \frac{1}{3}\zeta_0(\chi) + \frac{1}{3}\zeta_2(\chi) \cos(2\theta),
\end{equation}
where the radial functions $\zeta_0$ and $\zeta_2$ are solutions of the forced Bessel equations
\begin{eqnarray}
\hat{L}_m \zeta_m &=& J_1^2(\chi); \ \ m =0,2,\nonumber\\
\d{\zeta_m}{\chi}(\tilde{R}_1) &=& 0,
\end{eqnarray}
whereas
\begin{equation}
\hat{L}_m = \frac{1}{\chi}\d{}{\chi}\chi\d{}{\chi}+1-\frac{m^2}{\chi^2}
\end{equation}
is the Bessel operator of order $m$.

\begin{figure}[t]
\includegraphics[scale=0.45]{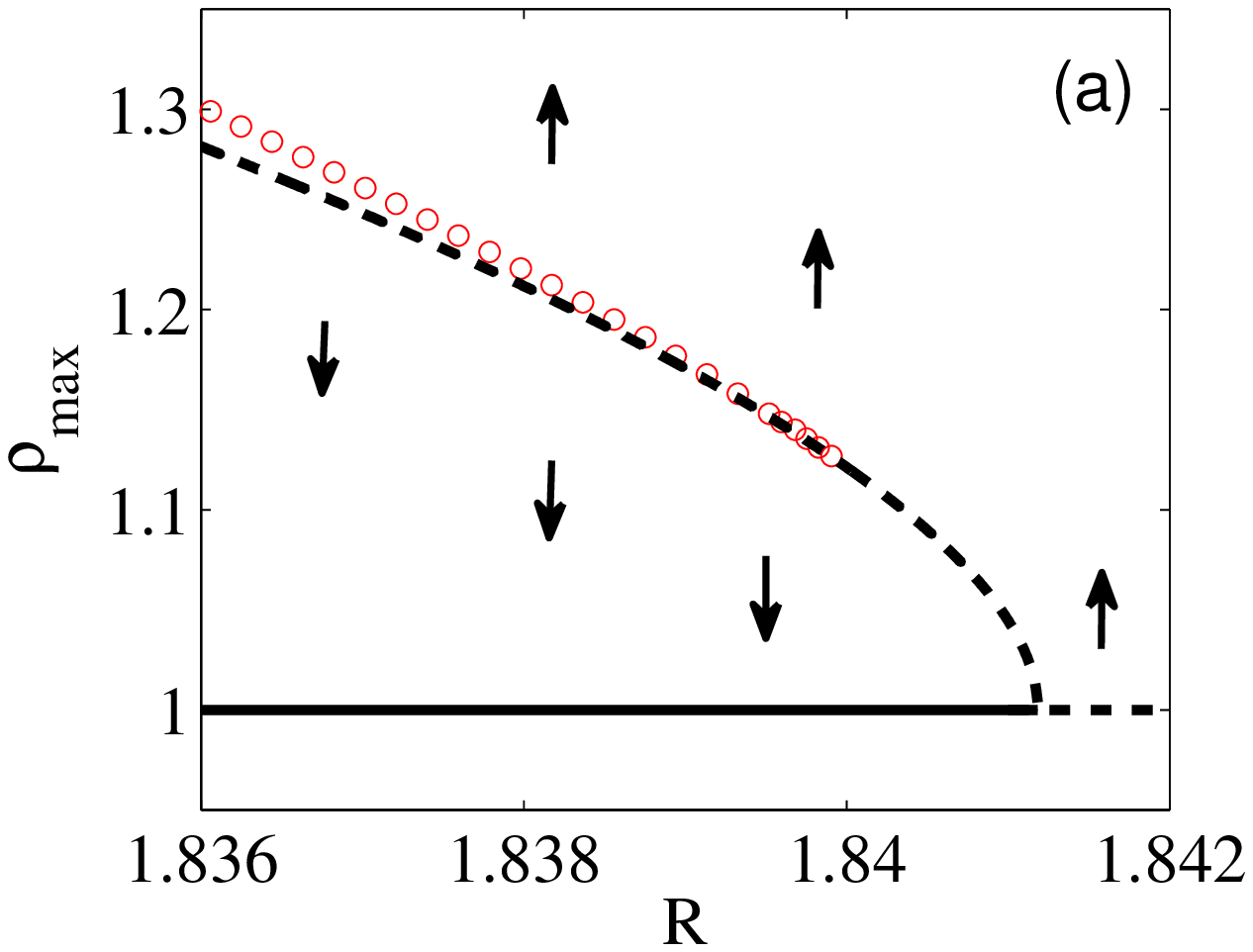}\\
\includegraphics[scale=0.45]{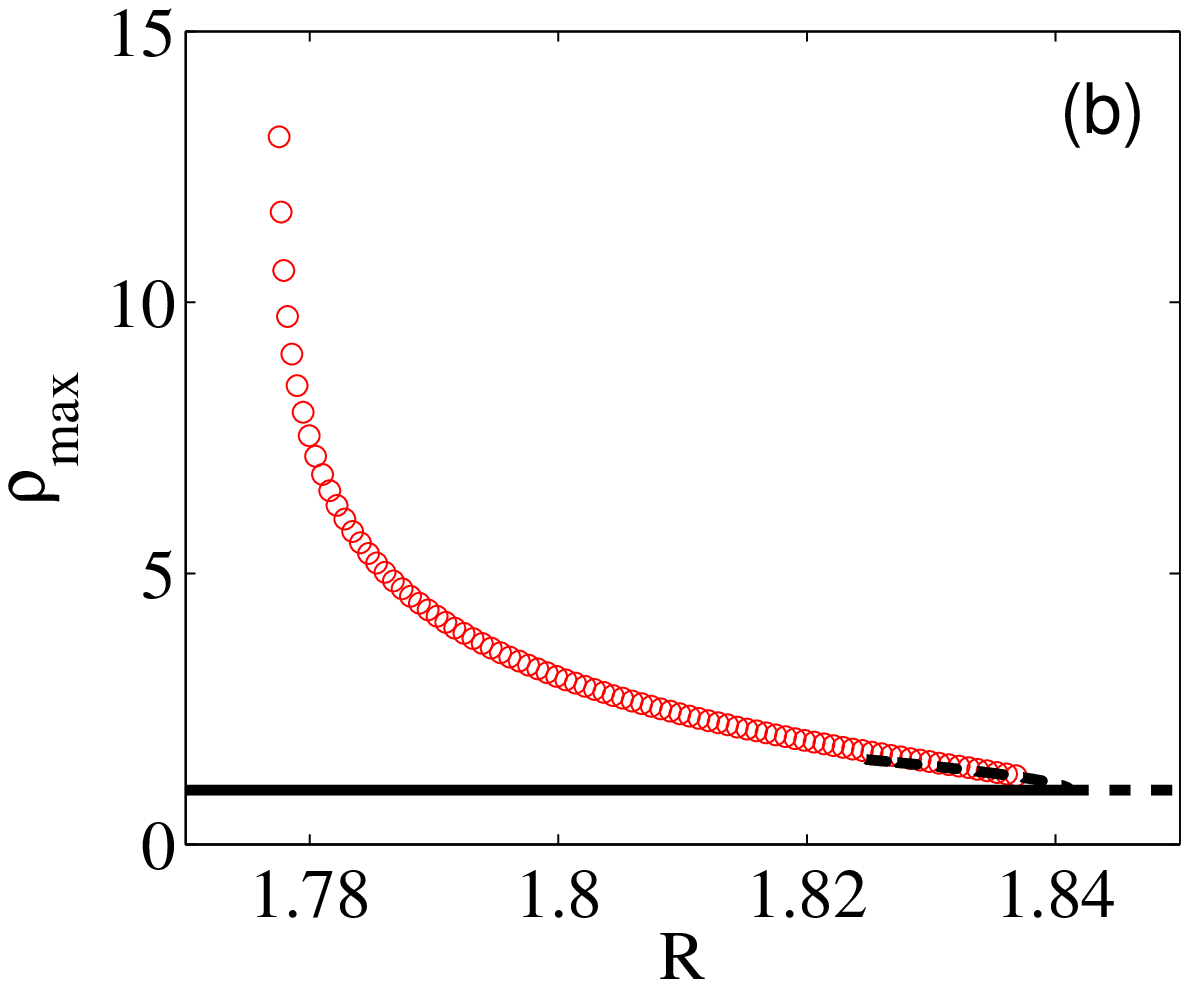}
\caption{(Color online.) Time-independent inhomogeneous cooling states  in a circular container exhibit a sub-critical pitchfork bifurcation. Shown is the gas density at the circumference $r=R$ versus the container radius $R$. The solid lines denote stable states, the dotted lines denote unstable states. The horizontal line $\rho_{max} = 1$ depicts the HCS. The dashed lines depict analytical solutions.  The circles show results of numerical integration of Eq.~(\ref{disk_problem_conv}). All the quantities are plotted in rescaled units. The arrows schematically show the direction of the evolution of the flow depending on the stability properties of the branches. Panel (b) is an extension of panel (a) to smaller container radii.} \label{m1_bif_fig}
\end{figure}

\begin{figure}[t]
\includegraphics[scale=0.5]{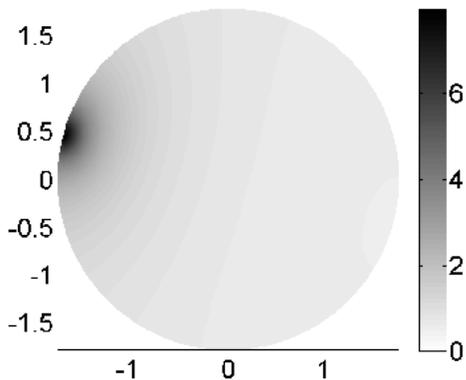}
\caption{The density map of the \textit{unstable} inhomogeneous steady-density state in a circular container of radius $R \simeq 1.7795 \simeq 0.9666\,R_1$.  This state, obtained by solving numerically Eq. (\ref{disk_problem_conv}), serves as a ``border" between those cooling flows which approach the HCS and those which do not.} \label{m1_inho_state}
\end{figure}

In the third order in $\epsilon$ we obtain
\begin{eqnarray}
\label{disk_problem_3rd}
\nabla^2\varphi_3+\varphi_3 &=& \frac{4}{3}\varphi_1\varphi_2-\frac{14}{27}\varphi_1^3-2k_2\varphi_1\nonumber\\
\pd{\varphi_3}{\chi}(\tilde{R}_1,\theta) &=& 0.
\end{eqnarray}
The solvability condition again demands to choose $k_2$ so that the right-hand side be orthogonal to $\varphi_1$. This is achieved by setting
\begin{equation}
k_2 = \frac{\int\int_D\left(\frac{2}{3} \varphi_1\varphi_2 -\frac{7}{27}\varphi_1^3\right)\varphi_1\chi d\chi d\theta}{\int\int_D\varphi_1^2\chi d\chi d\theta} = -0.02153\dots.
\end{equation}
Here the integration $\int\int_D \dots \chi d\chi d\theta$ is carried out over the circular container $D$ of radius $\tilde{R}_1$.
With this value of $k_2$ we can calculate the dependence of $\epsilon$ on $\delta \tilde{R} = \tilde{R} - \tilde{R}_1$:
\begin{equation}
\epsilon^2 \simeq \frac{k-1}{k_2}= \frac{1}{\tilde{R}_1 k_2}\delta \tilde{R}.
\end{equation}
As $k_2<0$,  $\delta \tilde{R}$ can only take negative values. That is, somewhat surprisingly, a time-independent solution describing an inhomogeneous density distribution is possible only in a container with radius \textit{smaller} than $\tilde{R}_1$.  Going back to the rescaled variables (that is, two transformations back), we find
\begin{eqnarray}
A^{-2} &=& 1 + \frac{\avg{J_1(\chi)^2}_\chi}{18}\,\epsilon^2,\nonumber\\
\delta R &\equiv & R - R_1 \simeq \left[1 + \frac{\avg{J_1(\chi)^2}_\chi}{18 k_2} \right] \delta \tilde{R}, \nonumber\\
\rho(r,\theta)& \simeq &1-\frac{2}{3}\epsilon \,J_1\left(\frac{R_1}{R} r\right)\cos(\theta).
\end{eqnarray}
Now we can determine the form of the bifurcation diagram in the (rescaled) original coordinates. Taking $\delta R=R-R_1$ to be the bifurcation parameter, and the density maximum $\rho_{max}$ to be the bifurcation amplitude, we obtain
\begin{equation}
\label{m1_bif}
\rho_{max} \simeq 1+3.1432\dots\sqrt{R_1-R},\;\;\; R_1-R \ll R_1.
\end{equation}
This is a subcritical bifurcation, where inhomogeneous solutions exist only at $R<R_1$. The time-independent inhomogeneous solutions form a continuous set
which can be obtained by rotating one of the solutions by an arbitrary angle around the center of the container. Figure~\ref{m1_bif_fig} presents the bifurcation diagram alongside with its extension obtained by numerical integration of Eq. (\ref{disk_problem_conv}) (using the \texttt{pdenonlin} algorithm of the  Matlab PDE toolbox) and transformation back to the (rescaled) original variables. The well-known properties of supercritical bifurcations  in 1D \cite{Crawford} make it possible to determine the stability of the branches shown in Fig.~\ref{m1_bif_fig} without further calculations.   The arrows in Fig. \ref{m1_bif_fig}  schematically show the direction of the evolution of the flow depending on whether the branch is stable or unstable. We found that the maximum density of the unstable states diverges at a finite subcritical container radius, $R \simeq 1.7795 \simeq 0.9666\,R_1$, see Fig.~\ref{m1_bif_fig}b. This suggests that this is a critical radius separating, in a proper space of parameters,  the flows to those which approach the HCS from those which do not. Figure~\ref{m1_inho_state} shows the numerically calculated density map of the ``border state" corresponding to this critical radius of the container.

One salient feature of the bifurcation diagram in Fig. \ref{m1_bif_fig} is that there are no stable time-independent inhomogeneous solutions at $R>R_1$. This result is in a striking contrast with that for the narrow-channel geometry, where the pitchfork bifurcation for the clustering mode is \textit{super}-critical, and where the energy loss and heat diffusion balance each other to produce stable time-independent inhomogeneous states \cite{MFV}. What is then the ultimate state of a freely cooling granular gas at $R>R_1$?

\subsection{Hydrodynamic simulations in a circular container}
\label{hydrocircular}

\begin{figure}[t]
\epsfclipon
\begin{tabular}{cc}
\includegraphics[scale=0.5]{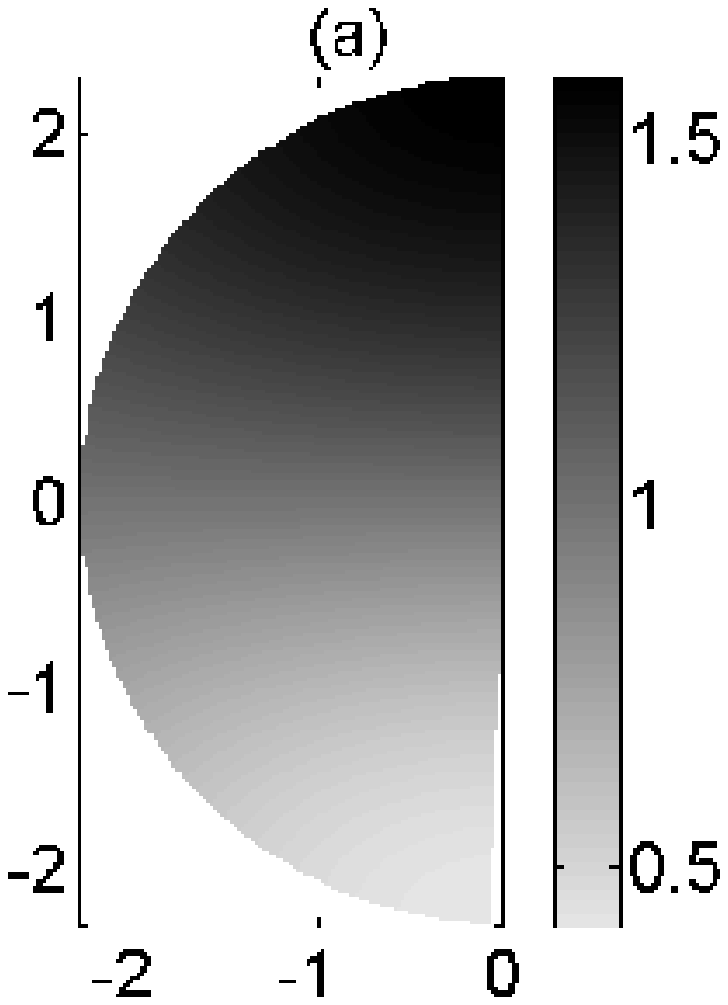} &
\includegraphics[scale=0.5]{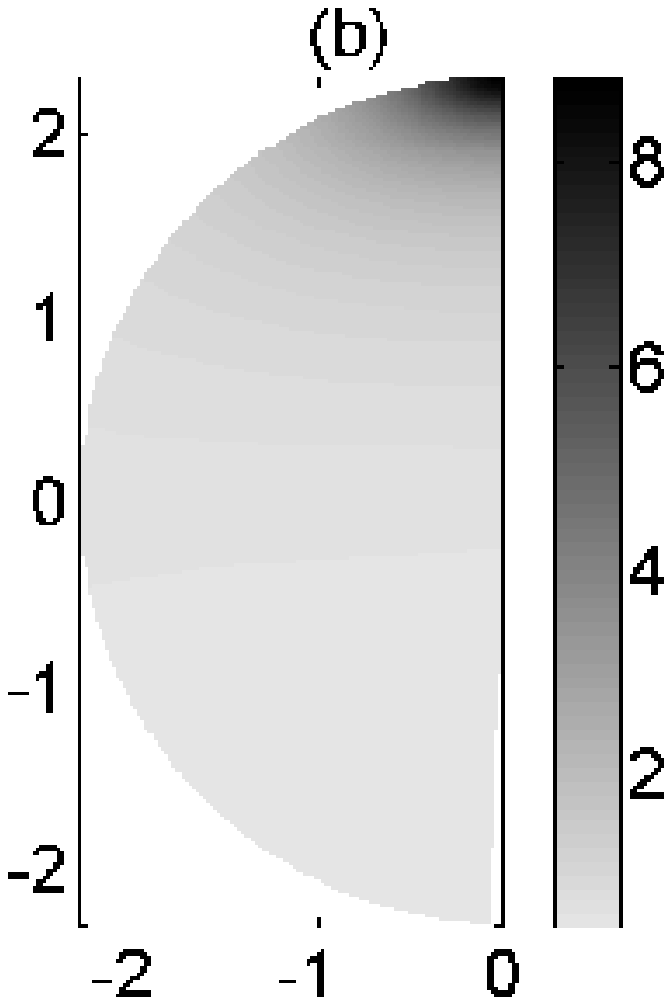}\\
 \includegraphics[scale=0.5]{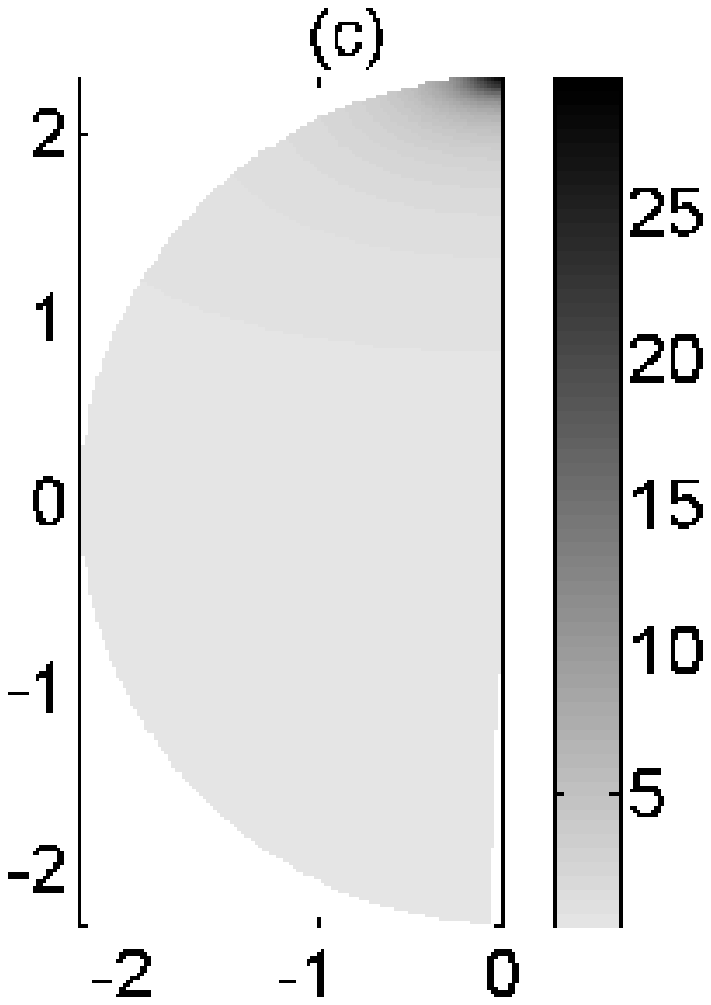} &
\includegraphics[scale=0.5]{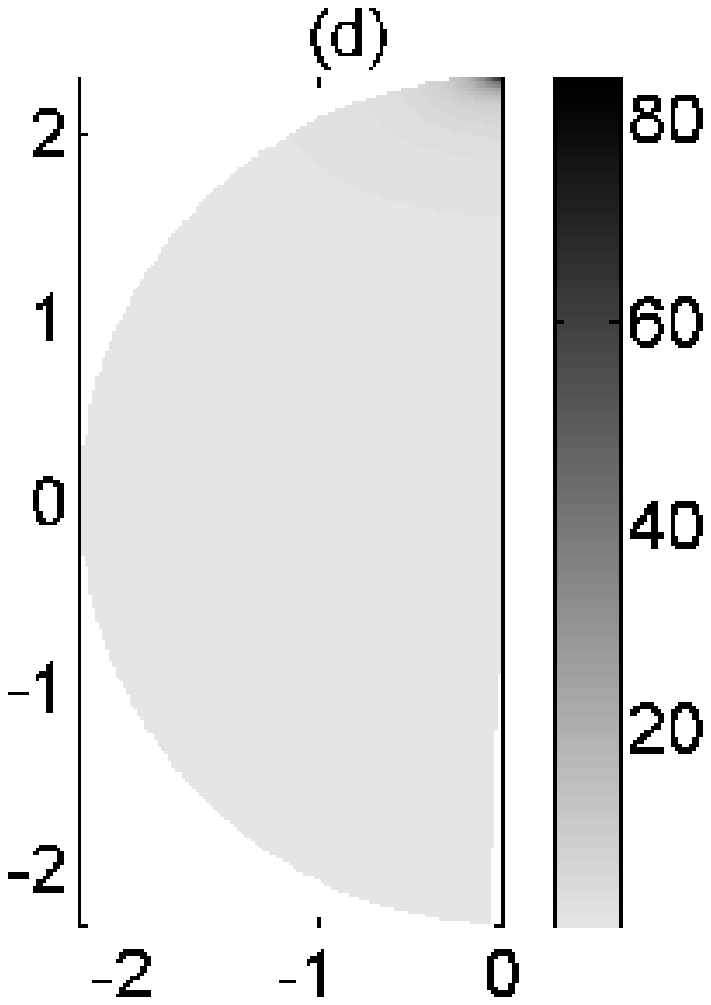} \\
\end{tabular}
\caption{The density history of a freely cooling granular gas in a circular container of rescaled radius $R=1.3R_1$ as obtained in a hydrodynamic simulation for initial conditions~(\ref{disk_initial}). One half of the container is shown. The (rescaled) times of the snapshots are $t= 1.7$ (a), $19.1$ (b), $32.12$ (c) and $39.45$ (d).} \label{m1}
\end{figure}

\begin{figure}[t]
\includegraphics[scale=0.5]{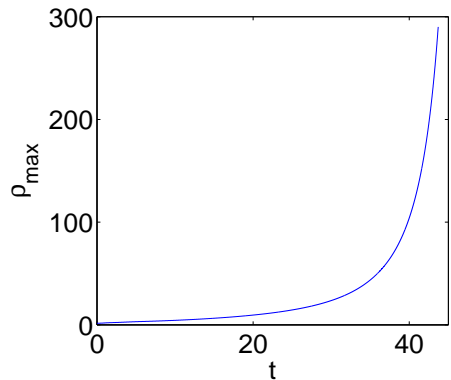}
\caption{The maximum gas density versus time corresponding to Fig. \ref{m1}.} \label{m1_blowup}
\end{figure}

Our first step in answering this question was performing a series of 2D hydrodynamic simulations in this geometry, see Appendix A for the computational details. After a proper rescaling the time-dependent hydrodynamic equations include only one dimensionless parameter, proportional to $1-\alpha^2$, see Ref. \cite{MFV} and Appendix A.  We chose the restitution coefficient $\alpha=0.94868\dots$ which corresponds to $\epsilon_1=\kappa_0\Lambda/2 = 1-\alpha^2 = 0.1$.  The hydrodynamic equations, subject to boundary conditions (\ref{temp_BC}) and (\ref{slip_BC}), were solved with the (zero-viscosity) Vulcan/2D code, see Appendix A. An Eulerian numerical scheme on a non-uniform polar mesh was used.  Here we present the results of a typical simulation, performed in (one half of a) circular container of rescaled radius $R = 2.3$. This value is larger than the rescaled critical radius $R_1 = 1.84\dots$ for the $m=1$ clustering mode, but smaller than the critical radii for the shear mode and for all other clustering modes in this geometry. The initial conditions were
\begin{eqnarray}
\label{disk_initial}
\rho(r,\theta ,t=0) &=& 1 + J_1(R_1 r /R)\cos(\theta),\nonumber\\
v(r,\theta , t=0) &=& 0,\nonumber\\
p(r,\theta , t=0) &=& 1.
\end{eqnarray}
The simulation results are illustrated in Figs. \ref{m1} and \ref{m1_blowup}.
Figure \ref{m1} shows density map snapshots at different (rescaled) times. One can see that the gas develops a density peak at the point $r=R, \theta = 0$.  Figure \ref{m1_blowup} shows the evolution of this density peak in time.  The numerical run was stopped when the (rescaled) maximum density reached about $300$, and a considerable amount of gas was already concentrated in a region comparable with the mesh size. One can see that, until this time, the density growth \textit{accelerates} in time. These results suggest that the gas is developing thermal collapse: a finite-time density blowup which, contrary to what happens in a narrow-channel flow \cite{MFV},  is not arrested by heat diffusion. To substantiate this hypothesis, we will now switch to a narrow-sector geometry, see Fig. \ref{Geometries}.

\section{Time-independent Inhomogeneous Cooling States in a narrow sector}
\label{cluster}
In a  narrow-sector-shaped container the heat diffusion suppresses any flow structure in the azimuthal direction. This setting, therefore, defines a purely radial  (1D) flow while retaining an essential 2D feature:  polar metrics.

As the (azimuthally-asymmetric) $m=1$ clustering modes are now suppressed, the first mode that goes unstable, as the sector radius $R$ is increased from zero, is the lowest $m=0$ clustering mode, see Sec. \ref{hydro}C. Let the sector radius be slightly larger than the critical radius $R_0=3.83171 l_\kappa$ for the lowest $m=0$ mode. The unstable $m=0$ mode will start to grow.  What is the long-time behavior of the gas in this geometry? To answer this question, we will first identify a time-independent inhomogeneous solution: a possible candidate for the description of a long-time behavior. We will show that the bifurcation of the HCS is trans-critical here, and determine its sign by performing a weakly nonlinear analysis.

\subsection{Bifurcation analysis}
Let the rescaled sector radius be slightly larger, or slightly smaller, than $R_0=3.83171\dots$. We again use the transformed equation (\ref{disk_trans}). Dropping the $\theta$-derivatives, we obtain the nonlinear eigenvalue problem
\begin{eqnarray}\label{sector_trans}
\frac{1}{3}\dd{\Phi}{\tilde{r}} + \frac{1}{3\tilde{r}}\d{\Phi}{\tilde{r}}+1-\Phi^{-1/3}=0, \nonumber \\
\d{\Phi}{\tilde{r}}(\tilde{r}=\tilde{R}) = 0,
\end{eqnarray}
the sector radius serving as the eigenvalue. As the HCS corresponds to $\Phi = 1$, we substitute $\Phi = 1 + \epsilon J_0(\tilde{r}) +\epsilon^2 \varphi_2(\tilde{r})$ in Eq.~(\ref{sector_trans}) and neglect terms of orders $\epsilon^3$ and higher. The resulting equation is a forced Bessel equation:
\begin{equation}\label{cluster_WNL}
\dd{\varphi_{2}}{\tilde{r}} + \frac{1}{\tilde{r}}\d{\varphi_{2}}{\tilde{r}}+\varphi_{2}=\frac{2}{3} J_{0}^2(\tilde{r})\,.
\end{equation}
This equation can be solved via variation of parameters, and we obtain
\begin{eqnarray}
\label{WNL_solution}
\Phi = 1+\epsilon J_{0}(\tilde{r}) - \frac{\pi}{3} \epsilon^2 J_{0} (\tilde{r})\int_{0}^{\tilde{r}} J_{0} (\tilde{r}')^2 Y_{0} (\tilde{r}^{\prime}) \tilde{r}^{\prime} d\tilde{r}^{\prime} \nonumber \\
+ \frac{\pi}{3} \epsilon^2 Y_{0} (\tilde{r})\int_{0}^{\tilde{r}} J_{0} (\tilde{r}^{\prime})^3 \tilde{r}^{\prime} d\tilde{r}^{\prime}\,,
\end{eqnarray}
where $Y_0(\dots)$ is the Bessel function of the second kind. This inhomogeneous solution must hold at $\tilde{R} \neq R_0$, so $\epsilon$ depends on $\tilde{R}$. We can seek $\tilde{R}=R_0+\delta R$, where $\delta R \ll R_0$.  Forcing the solution to obey the boundary condition
\begin{equation}
\d{\Phi}{\tilde{r}}(R_0 + \delta R)=0,
\end{equation}
we obtain, in the lowest order, a linear relation between $\epsilon$ and
$\delta R$:
\begin{equation}\label{m0_trans_bif}
\epsilon = - \frac{3J_{1}^{^{\prime}} (R_0)}{\pi Y_{1}(R_0) \int^{R_0}_{0} \tilde{r} J_{0}^3(\tilde{r}) d\tilde{r}} \delta R =2.223\dots\, \delta R\,.
\end{equation}
Returning to the rescaled variables $r$ and $\rho$, we obtain the density at the arc of the sector, $r=R$:
\begin{equation}\label{m0_bif}
\rho (r=R) = 1 - \frac{2}{3} J_0(R_{0}) \epsilon =1+0.5968\dots \,  (R-R_0) ,
\end{equation}
to first order in $R-R_0$.

\begin{figure}[ht]
\includegraphics[scale=0.5]{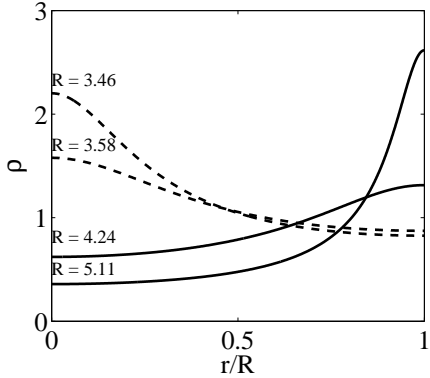}
\caption{Inhomogeneous cooling states below and above the bifurcation point $R=R_0$, obtained by numerical integration of Eq.~(\ref{sector_trans}) for different sector radii. The solid curves show stable solutions, the dashed curves show unstable ones. The quantities are plotted in rescaled units.} \label{m0_solutions}
\end{figure}

\begin{figure}[t]
\includegraphics[scale=0.45]{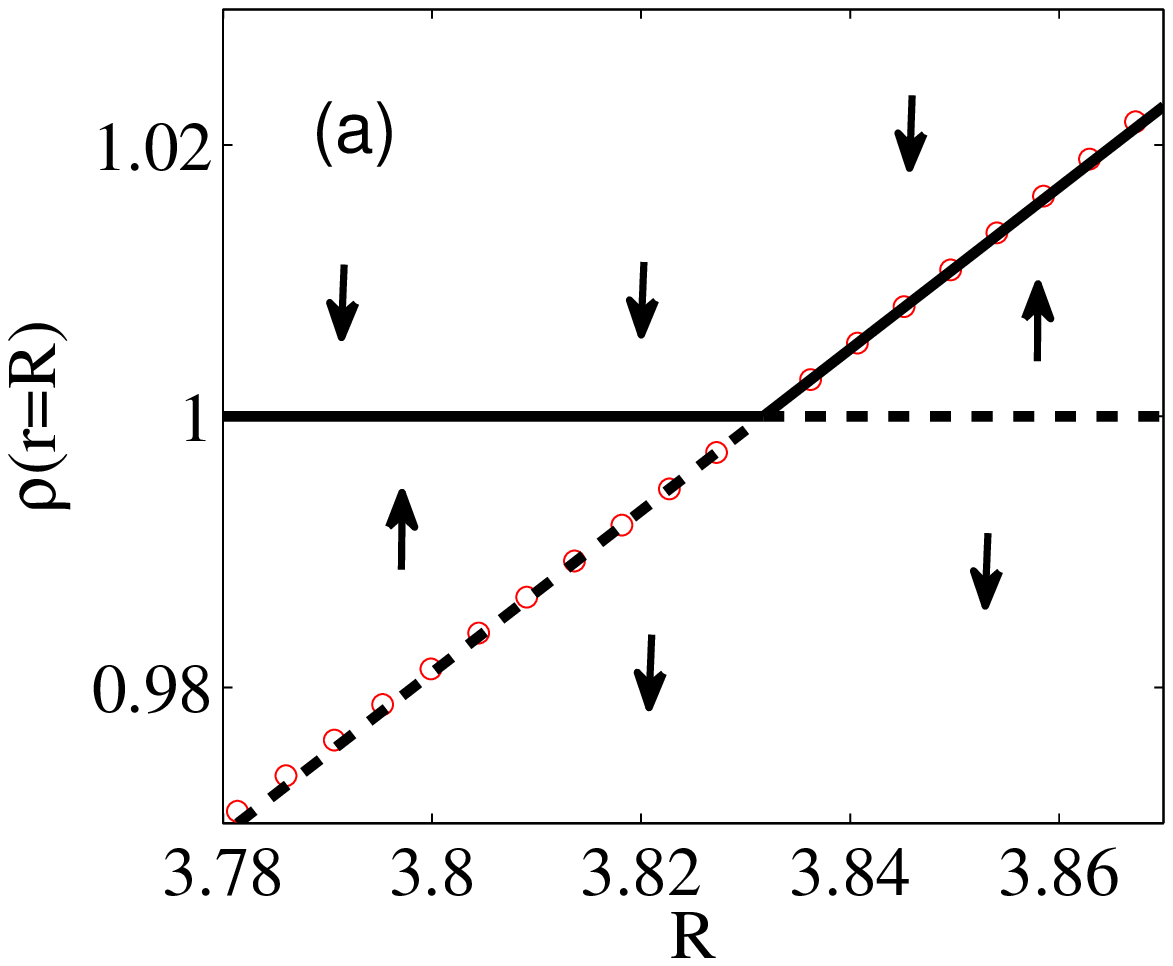}\\
\includegraphics[scale=0.45]{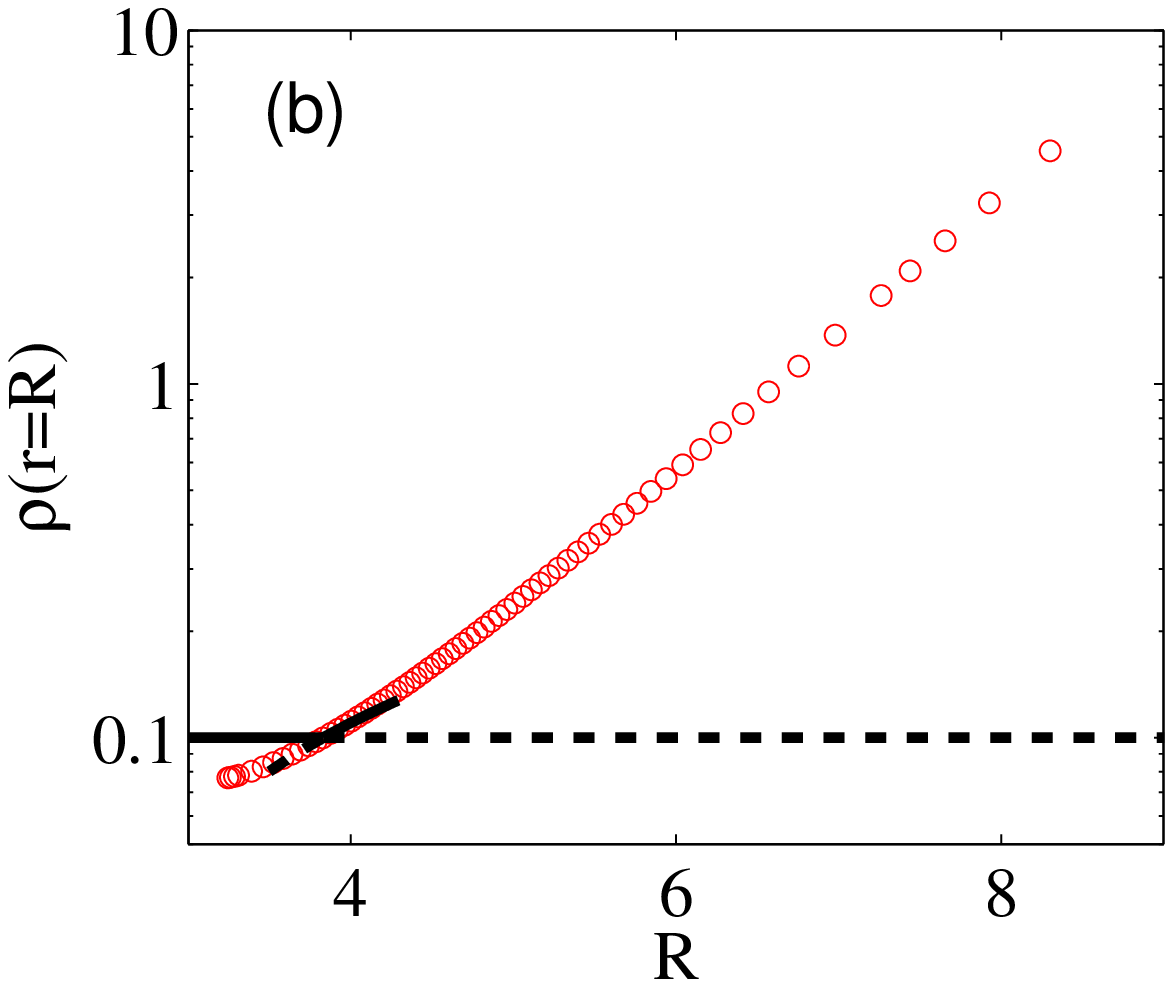}
\caption{(Color online.) Time-independent inhomogeneous cooling states  in a narrow-sector-shaped container exhibit a trans-critical bifurcation.  Shown is the gas density at the arc $r=R$ versus the sector radius $R$. The solid lines denote stable states, the dashed lines denote unstable states. The HCS is represented by the horizontal line $\rho(R) = 1$. The inclined straight line is the relation (\ref{m0_bif}). The results of numerical integration of Eq.~(\ref{sector_trans}) are shown as  circles. The arrows indicate the conjectured flow in a system's phase space. All the quantities are plotted in rescaled units.  Panel (b) is an extension of panel (a) to larger and smaller sector radii.} \label{bifurcationdiagram}
\end{figure}

It is clear from this calculation that, at least for system sizes close to $R_0$, the cooling gas has two possible steady-density states. The first is the HCS $\rho = 1 $. The other is an inhomogeneous state where, for  $R<R_0$, the density is peaked at the vertex of the sector, and for $R>R_0$ the density is peaked at the arc. Several examples of density profiles of these inhomogeneous states, obtained by numerical integration of Eq.~(\ref{sector_trans}), are shown in Fig. \ref{m0_solutions}.

Equation (\ref{m0_bif}) describes a trans-critical bifurcation \cite{Crawford} in which two states ``collide" and exchange their stability. The bifurcation parameter is $\delta R$. As $R$ increases, the HCS remains stable until $R=R_0$. Here the HCS ``surrenders" its stability to the inhomogeneous state. Correspondingly, the inhomogeneous state is unstable at $R<R_0$. The bifurcation diagram of the system is shown in Fig. \ref{bifurcationdiagram} a.

\begin{figure}[ht]
\begin{tabular}{cc}
\includegraphics[scale=0.5]{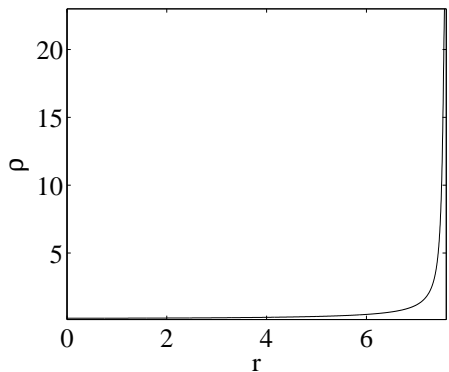}
\end{tabular}
\caption{A steady-state density profile of a freely cooling granular gas in a narrow-sector-shaped container of rescaled radius $R = 7.57$. The quantities are plotted in rescaled units.}\label{lstate}
\end{figure}

\begin{figure}[ht]
\begin{tabular}{cc}
\includegraphics[scale=0.5]{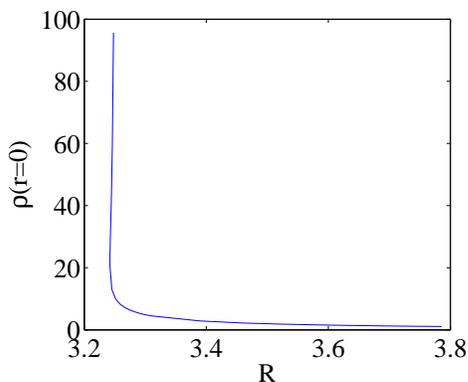}
\end{tabular}
\caption{The gas density at the vertex of a narrow sector versus the sector radius for \textit{unstable} inhomogeneous solutions at $R<R_0$. The quantities are plotted in rescaled units.}\label{m0_lower_branch_1}
\end{figure}

The bifurcation diagram can be extended beyond the weakly-nonlinear regime by numerical integration of Eq.~(\ref{sector_trans}).  Figure \ref{bifurcationdiagram} b shows the computed stable branch of the inhomogeneous solutions at $R>R_0$, and the unstable
branch at $R<R_0$. Note the exponential dependence, at large radii, of the maximum density on the sector radius. It has the same origin as the exponential system-size dependence of the maximum density of the inhomogeneous cooling state in a long narrow channel \cite{MFV}. Our numerics show that the unstable branch ends at $R \simeq 3.25$, where $\rho(R) \simeq 0.77$. That is, there are no time-independent inhomogeneous solutions for the density at smaller radii. Figure \ref{m0_lower_branch_1} shows, for the unstable branch, how the density at the vertex, $\rho(r=0)$, grows when $R$ goes down.

When increasing the sector radius, the density profile becomes more and more peaked at $r=R$, so one can call these states localized \cite{Fouxon3}.  In Fig. \ref{lstate} one can see the radial density profile for $R = 7.57$.

The bifurcation diagram in Fig.~\ref{bifurcationdiagram} suggests a ``phase diagram" of the cooling gas in the sector geometry. Let us denote the time-independent inhomogeneous solution of Eq.~(\ref{ICS_density}) in a sector of radius $R$ by $\rho_I(r;R)$. Consider a continuous one-parameter family of initial conditions such that the density profile is a linear combination of the inhomogeneous solution $\rho_I(r;R)$ and the HCS $\rho=1$:
\begin{eqnarray}
\label{sector_initial}
\rho(r,t=0) &=& C \rho_I(r;R)+1-C,\nonumber\\
v(r,t=0) &=& 0, \nonumber \\
p(r,t=0) &=& 1.
\end{eqnarray}
The constants are chosen so that $\langle \rho\rangle=1$. Let us identify an arbitrary point $[R,\rho]$ of the plane of Fig. \ref{bifurcationdiagram} with this initial condition. Then the constant $C$
is determined by $\rho(R) = C \rho_I(R;R)+1-C$, so
\begin{equation}
C=\frac{\rho_I(R;R)-1}{\rho(R)-1}.
\end{equation}
For initial conditions located above the unstable branches of Fig. \ref{bifurcationdiagram}, the gas will approach the corresponding stable state: the HCS at $R<R_0$, and the inhomogeneous cooling state at $R>R_0$. There are no stable states, however, when starting \textit{below} the unstable branches, see Fig. \ref{bifurcationdiagram}. This suggests that, as in the circular container, the cooling gas develops thermal collapse. In the following section we will present analytical and numerical evidence supporting this conjecture.

\section{Thermal collapse in 2D}
\label{collapse}
\subsection{Reduced ideal hydrodynamics close to singularity}
\label{isobaric}
That a freely cooling dilute granular gas exhibits thermal collapse, in any dimension, within the framework of \textit{ideal} granular hydrodynamics, is by now well established \cite{Fouxon1,Fouxon2,PuglisiA,MFV,Fouxon3}.  In this subsection we will derive a set of reduced ideal hydrodynamic equations which describe thermal collapse in 2D. In the next subsection we will obtain scaling laws, in time and in space, which characterize the thermal collapse. Then we will check whether heat diffusion and other processes, that we will neglect, can become relevant near the singularity.

Assuming that thermal collapse does happen, and that heat diffusion and viscosity are irrelevant in the vicinity (both in time, and in space) of the density singularity, one can considerably simplify the hydrodynamic equations Eqs.~(\ref{continuity}), (\ref{NS}) and (\ref{heatp}). In the light of results of Refs. \cite{Fouxon1,Fouxon2,PuglisiA,MFV,Fouxon3}, an important additional simplification comes from the assumption that the gas pressure in the collapse region is approximately uniform. In a formal language, see \textit{e.g.} Ref. \cite{Glasner}, we eliminate the acoustic modes by postulating a perturbation expansion for the gas pressure,
\begin{equation}
\label{p_series}
p(\mathbf{r},t) = p^{(0)}(t)+p^{(1)}(\mathbf{r},t)+\dots,\;\;\;p^{(1)} \ll p^{(0)},
\end{equation}
limit ourselves to the leading-order approximation, $p=p^{(0)}(t)$, and discard the momentum equation (\ref{NS}). Finally, we neglect both terms in the left hand side of Eq.~(\ref{heatp}), as they remain bounded at the density singularity, and keep only terms which blow up at the singularity. We arrive at the following reduced equations:
\begin{eqnarray}
\pd{\rho}{t}+\nabla \cdot (\rho\mathbf{v}) &=&0, \label{isobaric_singularity} \\
\gamma p_s^{-1/2}\nabla\cdot\mathbf{v} &=& -\Lambda \rho^{1/2}\,, \label{isobaric_singularity1}
\end{eqnarray}
where $p_s$ is the gas pressure at singularity. Equation (\ref{isobaric_singularity1}) describes a balance between the adiabatic compression heating and the inelastic cooling.

Set of equations (\ref{isobaric_singularity}) and (\ref{isobaric_singularity1}) is not closed, as there are only two scalar equations for the density and the two velocity components. To obtain an additional equation, for the velocity, one should go to the subleading order in Eq.~(\ref{p_series}), as it was done in Ref. \cite{Glasner} in the case of a 2D gas flow driven by heat diffusion. Instead, we will make another simplifying assumption: that the gas flow near thermal collapse is (at least locally) azimuthally symmetric. Note that, if thermal collapse develops at the vertex of a narrow-sector-shaped container, this assumption is satisfied trivially. For an azimuthally-symmetric flow Eqs.~(\ref{isobaric_singularity}) and (\ref{isobaric_singularity1}) make a closed set.

\subsection{Scaling laws of thermal collapse}
\label{scaling}

Let us rescale $t$ by
$t_s = 2\gamma/(\Lambda p_s^{1/2} \rho_0^{1/2})$, $\rho$ by the average gas density $\rho_0$, $r$ by $l_\kappa$ and $v$ by $l_\kappa/t_s$. Then, for a $d$-dimensional thermal collapse (azimuthally symmetric in 2D or spherically symmetric in 3D) Eq.~(\ref{isobaric_singularity1}) acquires a parameter-free form:
\begin{equation}
\label{singular_heat_isotropic}
\frac{1}{r^{d-1}}\pd{}{r}\left(r^{d-1} v\right) = - 2 \rho^{1/2}\,,
\end{equation}
where $r$ is a local radial coordinate with the origin at the point of developing singularity.
Substituting Eq.~(\ref{singular_heat_isotropic}) in Eq.~(\ref{isobaric_singularity}) we obtain
\begin{equation}
\label{Lag_start}
\pd{\rho}{t} + v\pd{\rho}{r} = 2 \rho^{3/2}.
\end{equation}
This equation becomes very simple in Lagrangian coordinates which transform the full time derivative on the left-hand-side into a simple time derivative. We can define a Lagrangian mass coordinate
\begin{equation}
m(r,t) = \int_0^r \rho(r^{\prime},t) {r^{\prime}}^{d-1} dr^{\prime},
\end{equation}
$r^{d-1}$ being the metric in $d$ dimensions. Up to a $d$-dependent factor, $m(r,t)$ is the mass of the gas contained in a $d$-dimensional sphere of radius $r$ around the point of developing singularity. After the transformation Eq.~(\ref{Lag_start}) becomes remarkably simple:
\begin{equation}
\pd{w}{t}\big|_{m=const} = -1,
\end{equation}
where $w(m,t) = \rho^{-1/2}(m,t)$. The solution is
\begin{equation}\label{sol}
w(m,t) = -t + w_0 (m),
\end{equation}
where $w_0(m)$ is the initial condition (with a minimum at $m=0$) which should be chosen at time when the reduced set of equations (\ref{isobaric_singularity}) and (\ref{isobaric_singularity1}) already holds. The time of singularity is therefore
$t_* = w_0(0)$. The scaling laws near collapse are determined by the behavior of the function $w_0(m)$ at small $m$. To elucidate this behavior, let us return to the density $\rho$ in the original Eulerian coordinates. Here, a generic density profile  near the singularity has a quadratic dependence on $r$:
\begin{equation}
\rho_0(r) \simeq a - b r^2,
\end{equation}
where $a>0$ and $b>0$. Using this local profile we can calculate,  for this ``initial condition", the local asymptote $m(r)$ near $r=0$:
\begin{equation}
m(r) = \int_0^r \rho_0(r^{\prime}) {r^{\prime}}^{d-1} dr^{\prime}\simeq \frac{a}{d} \, r^d,
\end{equation}
in the leading order in $r$.
Therefore,
\begin{equation}
\rho_0(m) \simeq a - b \left(\frac{d}{a}\right)^{2/d} \,m^{2/d}.
\end{equation}
This follows
\begin{equation}
w_0(m) \simeq a^{-1/2} + c \,m^{2/d},
\end{equation}
where $c>0$ is a constant. Now we see that the density blowup occurs at $t_* =  a^{-1/2}$. Now we return to Eq.~(\ref{sol}) and observe that, sufficiently close to the singularity point, the solution is self-similar:
\begin{equation}\label{w}
    w(m,t) \simeq t_* - t + c \,m^{2/d}=\tau\, W (m/ \tau^{d/2})\,,
\end{equation}
where $\tau =t_* - t$ and $W(z) = 1+c\, z^{2/d}$. Transforming back to the Eulerian coordinate, we obtain
\begin{eqnarray}
\label{r}
r^d &=& d\,\int_0^m \frac{dm^{\prime}}{\rho(m^{\prime},t)} = d\,\int_0^m dm^{\prime} \,\tau^2\,W^2\left(\frac{m^{\prime}}{\tau^{d/2}}\right) \nonumber \\  &=& \tau^{2+d/2} {\cal F} \left(\frac{m}{\tau^{d/2}}\right)\,,
\end{eqnarray}
where ${\cal F}(\dots)$ is a shape function. Equations (\ref{w}) and (\ref{r}) yield the dynamical length scale of the ideal singularity
\begin{equation}
\label{dynlenscale}
\ell\sim (t_*-t)^{\beta}, \;\;\; \beta=1/2+2/d.
\end{equation}
For $d=1$ (the narrow-channel geometry) we obtain $\beta=5/2$, in agreement with the result of Fouxon \textit{et al.} \cite{Fouxon1,Fouxon2}, whereas for $d=2$ and $3$ we obtain  new exponents $\beta=3/2$ and $7/6$, respectively.
The density asymptote close to the singularity can be written as
\begin{equation}
\label{self_similar_density}
\rho (r,t)= \tau^{-2}\Psi(\xi)\,,
\end{equation}
where $\xi = r/\tau^\beta$ is the similarity variable, and $\Psi$ is a shape function such that $\Psi(0)=1$. The density blows up at $r=0$ as $\rho = (t_*-t)^{-2}$ independently of dimension \cite{Fouxon3}.

At large distances, $r \gg \ell$ (but still much smaller than an external length scale of the flow) the density exhibits a time-independent power-law tail. Indeed, at $m\gg \tau^{d/2}$ asymptote~(\ref{w}) becomes $w(m,t)\simeq c \,m^{2/d}$. Then, using Eq.~(\ref{r}), we obtain $r^d \sim m^{1+4/d}$. As a result,
\begin{equation}
m \sim r^{\frac{d^2}{d+4}}
\end{equation}
and
\begin{equation}
\rho (r \gg \ell) \sim r^{\eta},\;\;\eta=-\frac{4d}{4+d} = -\frac{2}{\beta}.
\end{equation}
For $d=1$ this yields $\eta=-4/5$, in agreement with Refs.  \cite{Fouxon1,Fouxon2}, whereas for $d=2$ and $3$ one obtains  new exponents $\eta=-4/3$ and $-12/7$, respectively.

To find the scaling relations for the gas velocity we integrate Eq.~(\ref{singular_heat_isotropic})
\begin{equation}
\label{singular_velocity}
 v(r,t) = -  \frac{2}{r^{d-1}} \int_0^r \rho^{1/2} {r'}^{d-1}dr'\,,
\end{equation}
and change to the self-similar coordinate $\xi$:
\begin{equation}
\label{velocity_scaling}
v(r,t) = -\frac{2 \tau^{\beta-1}}{\xi^{d-1}} \int_0^\xi \Psi^{1/2}(\xi^{\prime}) \xi^{\prime d-1}\,d \xi^{\prime} \equiv - \tau^{\beta-1} V(\xi).
\end{equation}
When $\xi\gg 1$, the velocity field exhibits a time-independent power-law tail, $V\sim - \xi^{\mu}$, where $\mu=1-1/\beta$.  Again, for $d=1$ we recover the previous result $\mu=3/5$ \cite{Fouxon1,Fouxon2}. In 2D  $\mu= 1/3$, in 3D $\mu= 1/7$.

\subsection{Can heat diffusion arrest thermal collapse?}
\label{Implications}
Does heat diffusion, unaccounted for in ideal description, arrest the  collapse? In 1D the answer to this question is affirmative \cite{MFV}. What happens in higher dimensions?  Using the scaling relations in the vicinity of thermal collapse, that we have just found, we can estimate the ratio of the heat diffusion term to the cooling term. In the original units of Eqs. (\ref{continuity})-(\ref{heatp}) we obtain:
\begin{eqnarray}
\label{HD_vs_Cooling_D}
\abs{\frac{\kappa_0\nabla\cdot[\sqrt{p/\rho}\,\,\nabla (p/\rho)]}{\Lambda\rho^{1/2} p^{3/2}}} &\sim& \left(\frac{\rho}{\rho_0}\frac{\ell}{l_\kappa}\right)^{-2}\nonumber \\ &\sim& \left(\frac{t_*-t}{t_s}\right)^{3-\frac{4}{d}}.
\end{eqnarray}
One can see that the case of  $d=1$ is dramatically different from the cases of $d=2,3$. For $d=1$  heat diffusion cannot be neglected when time is sufficiently close to the $t_*$. As the heat diffusion is a stabilizing factor, one can argue that, in this case, it may arrest thermal collapse. Indeed, it was shown in Ref. \cite{MFV} that, in the narrow-channel geometry, heat diffusion ultimately balances the unstable density growth, and a time-independent density profile sets in. On the contrary, for $d=2$ or $3$ the heat diffusion remains irrelevant and cannot arrest the density blowup.

Estimate (\ref{HD_vs_Cooling_D}) implies that, in a narrow-sector-shaped container, thermal collapse is only possible exactly at the vertex and not in any other location.  Indeed, let $r=r_s>0$ be such another location. The differential of the local Lagrangian coordinate, measured from $r=r_s$,  is proportional to
$dr$. This setting is essentially 1D, and so heat diffusion will arrest the collapse  \cite{MFV}. On the contrary, the differential of the Lagrangian coordinate, measured from $r=0$,  is proportional to
$r dr$. That is, it is essentially 2D, and the scaling laws of the ideal singularity persist in spite of the heat diffusion.

In Appendix B we checked the consistency of other assumptions behind the reduced ideal set of equations (\ref{isobaric_singularity}) and (\ref{isobaric_singularity1}). As it turns out, one assumption -- about the (almost) uniform pressure profile -- breaks down at $t$ very close to $t_s$.
We will briefly discuss this issue in section \ref{discussion}.

\subsection{Hydrodynamic simulations in a narrow sector}
\label{hydro2}

\begin{figure}[ht]
\includegraphics[scale=0.5]{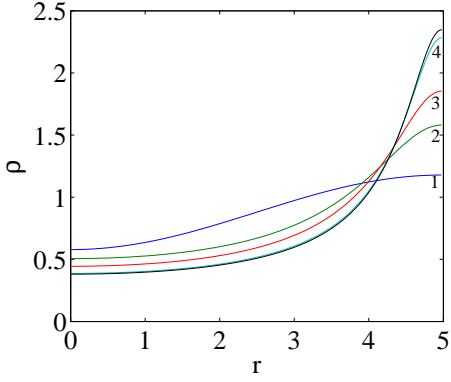}
\caption{(Color online.) The density profile $\rho(r,t)$  approaches a time-independent inhomogeneous state in a sector of radius $R = 1.3 R_0$. The profiles were obtained in a hydrodynamic simulation with initial conditions (\ref{sector_initial_static}) at times $t = 0$ (line 1),
$500$ (line 2), $2\cdot 10^4$ (line 3) and $t=10^9$ (line 4). The unnumbered line shows the stationary profile obtained from the solution of Eq.~(\ref{sector_trans}). The density $\rho$ is measured in units of $\rho_0$, and $r$ is measured in units of $l_\kappa$.}\label{m0_evolution}
\end{figure}

To test our theoretical predictions for a narrow-sector geometry, we solved numerically the rescaled hydrodynamic equations, see Appendix A. We assumed that the hydrodynamic fields depend only on the radial coordinate, whereas $v_{\theta}=0$, and used the zero-viscosity Vulcan/1D code, see Appendix A. Here we will describe two different simulations for $R = 1.3R_0$ (above the critical radius for the lowest $m=0$ mode). The restitution coefficient $\alpha=0.994987 \dots$ was chosen to satisfy the relation $\kappa_0\Lambda/2 = 1-\alpha^2 = 0.01$, whereas $\gamma$ was taken to be $2$. The first initial condition describes an isobaric configuration with a density peak at $r=R$:
\begin{eqnarray}
\label{sector_initial_static}
\rho(r,t=0) &=& 1 - 0.3\left[\frac{4}{\pi^2}+\cos\left(\frac{\pi r}{R}\right)\right],\nonumber\\
v(r,t=0) &=& 0, \nonumber \\
p(r,t=0) &=& 1. \\
\nonumber
\end{eqnarray}
Here we should expect, see Fig. \ref{bifurcationdiagram}, that the density profile will approach the corresponding time-independent inhomogeneous solution that we found in section \ref{cluster}. This is indeed what the simulated dynamics shows, see Fig. \ref{m0_evolution}.

\begin{figure}[ht]
\begin{tabular}{cc}
\includegraphics[scale=0.5]{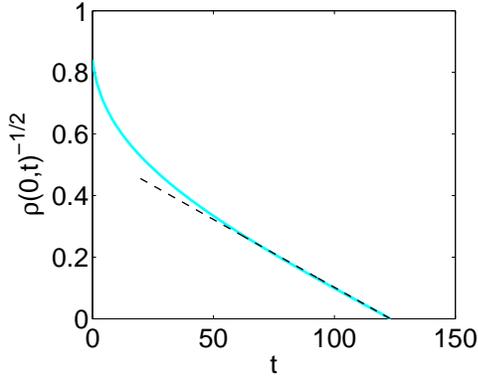}
\end{tabular}
\caption{(Color online.) Inverse square root of the gas density at the vertex of a narrow sector, versus time, obtained in a hydrodynamic simulation with initial conditions (\ref{sector_blowup_above}). The density blows up as $\sim (t_*-t)^{-2}$. The dashed straight line is a guide for the eye.
}\label{m0_invsqrtrho}
\end{figure}

\begin{figure}[ht]
\begin{tabular}{cc}
\includegraphics[scale=0.5]{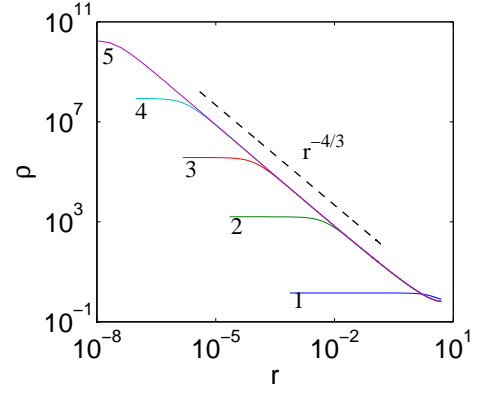}
\end{tabular}
\caption{(Color online.) Development of a density blowup at the vertex of a narrow sector of radius $R=1.3R_0$ in a hydrodynamic simulation with initial conditions (\ref{sector_blowup_above}). Shown are the density profiles at times $t = 0$ (line 1), $117.35$ (line 2),  $122.81$ (line 3),  $123.16$ (line 4) and  $123.18$ (line 5) in units of the cooling time $t_c$. The density is plotted in the units of the average density $\rho_0$. The time-independent part of the observed profiles exhibits a power law with exponent $-4/3$ predicted by our theory.
}\label{m0_blowup_above}
\end{figure}

The second initial condition differs from the first in that the initial density is now peaked at the vertex $r=0$:
\begin{eqnarray}
\rho(r,t=0) &=& 1 + 0.3\left[\frac{4}{\pi^2}+\cos\left(\frac{\pi r}{R}\right)\right]\nonumber\\
v(r,t=0) &=& 0, \nonumber\\
p(r,t=0) &=& 1. \label{sector_blowup_above}
\end{eqnarray}
The bifurcation diagram, see Fig. \ref{bifurcationdiagram}, suggests  that the gas will develop thermal collapse. The simulated history of the gas density at the vertex is depicted in Fig. \ref{m0_invsqrtrho}. Shown is the inverse square root of the density versus time. This time dependence is in excellent agreement, until the latest simulation times we could reach, with the finite-time density blowup $\rho(r=0,t)\sim(t_*-t)^{-2}$. Next we turn to the spatial behavior of the evolving density profiles. Figure \ref{m0_blowup_above} shows, in a log-log scale, several density profiles at different times. It is evident that, as the gas
density blows up at the vertex, the density profile develops a time-independent power-law tail with exponent $-4/3$ as predicted by our reduced ideal theory.

\begin{figure}[ht]
\includegraphics[scale=0.5]{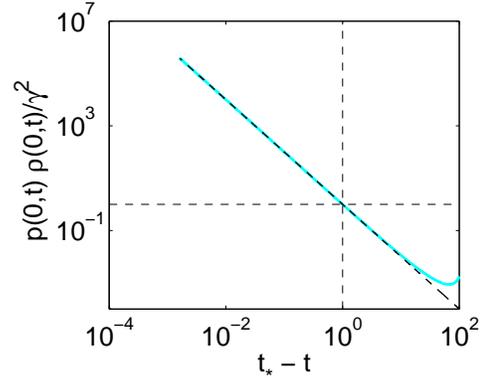}
\caption{(Color online.) Shown is $\gamma p(0,t)^{-1/2}\rho(0,t)^{-1/2}$ versus $t_*-t$ when approaching the density blowup for the hydrodynamic simulation with initial conditions (\ref{sector_blowup_above}). The time of the singularity $t_*=123.19$ was the only fitting parameter. The dashed line shows the function $(t_*-t)^{-2}$. The time is measured in units of $t_c$.}\label{m0_coefficient}
\end{figure}

\begin{figure}[ht]
\includegraphics[scale=0.5]{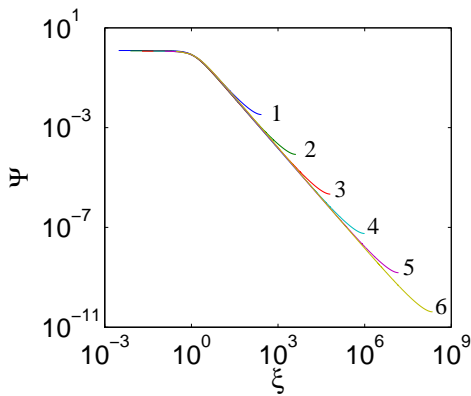}
\caption{(Color online.) Emergence of the self-similar shape function of the density for the hydrodynamic simulation with initial conditions (\ref{sector_blowup_above}). The times are $108.23$ (line 1), $120.8$ (line 2), $122.8$ (line 3), $123.12$ (line 4), $123.18$ (line 5) and $123.18$ (line 6), measured in units of $t_c$.  $p_s \simeq 7.765 \times 10^{-5}$.}\label{m0_collapse_collapse}
\end{figure}

\begin{figure}[ht]
\includegraphics[scale=0.5]{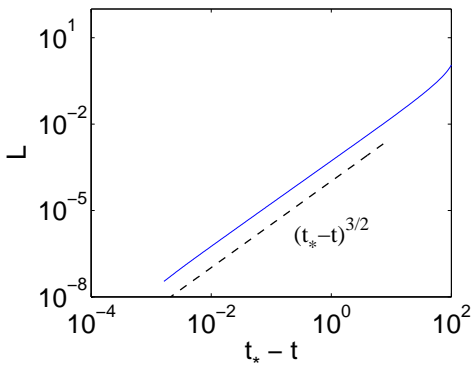}
\caption{(Color online.) The dynamical length scale of thermal collapse versus the time to singularity for the hydrodynamic simulation with initial conditions (\ref{sector_blowup_above}). The dashed line is shown to guide the eye.}\label{m0_dynscale}
\end{figure}

A more detailed check of the density blowup addresses the \textit{coefficient} of the scaling relation at $r=0$ as predicted by the ideal theory.  $\rho(0,t)$ should behave, close to blowup, as
\begin{equation}
\label{limit_density}
\rho(0,t) \to \frac{\gamma^2}{p_s (t_* - t)^2}\,,
\end{equation}
where the time is measured in units of $t_c$, the pressure in units of  the initial pressure $p_0$, and the density in units of $\rho_0$.
Using data from the simulation for the initial conditions (\ref{sector_blowup_above}), we plotted the quantity $\gamma p(0,t)^{-1/2}\rho(0,t)^{-1/2}$ versus time.  This  plot is shown, in a log-log scale, in Fig. \ref{m0_coefficient}. One can see that the curve  exhibits a power law with exponent $-2$ and passes through the point $(1,1)$, showing excellent agreement with Eq.~(\ref{limit_density}) in both the power law and the coefficient.
Now we proceed to verify the self-similarity of the density profiles at different times. Figures \ref{m0_collapse_collapse} depicts the shape function $\Psi(\xi)$ introduced in Eq.~(\ref{self_similar_density}). Notably, the validity range of the self-similar asymptote expands with time.
We also verified the power-law behavior in time of the dynamic length scale $\ell(t)$. As a measure of $\ell(t)$ in the simulations we defined the radial coordinate $L(t)$ where the gas density was equal to one half of the density at $r=0$. Figure \ref{m0_dynscale} shows the time dependence of $L(t)$, and excellent agreement with our analytical prediction in 2D, $\ell(t)\sim L(t) \sim \tau^{3/2}$ is observed.

\begin{figure}[ht]
\includegraphics[scale=0.5]{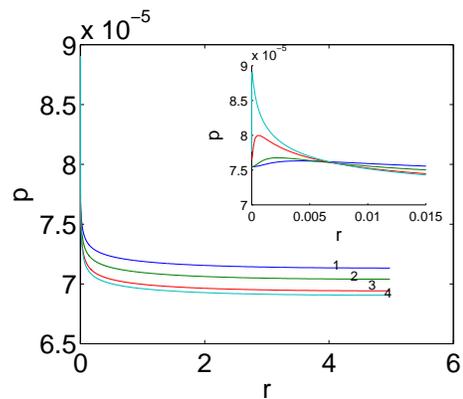}
\caption{(Color online.) The pressure profiles at $t=121.14$ (line 1), $121.96$ (line 2), $122.84$ (line 3) and $123.17$ (line 4) for the hydrodynamic simulation with initial conditions (\ref{sector_blowup_above}). The inset shows a close-up of these profiles near $r=0$.
As the blowup time approaches, the pressure profile steepens near $r=0$. The steepening is very modest, however, compared with the giant density growth at these times. Notice also that the characteristic length scale of steepening is larger by more than two orders of magnitude than the dynamic length scale $\ell(t)$ of thermal collapse.}\label{m0_pressure}
\end{figure}

Finally, most of our analytical results were obtained on the assumption that the gas pressure in the vicinity of thermal collapse remains uniform. We tested this assumption directly by following the simulated pressure profiles at different times. Figure \ref{m0_pressure} shows several pressure profiles in the whole sector, and a closeup in a vicinity of $r=0$.
One can see that, as thermal collapse progresses, the pressure remains close to uniform in the external region but exhibits some steepening near $r=0$. As the gas density varies by several orders of magnitude during these times, the spatial and temporal variations of the pressure are insignificant.   At the end of the simulation, the spatial length scale of the pressure is still much larger (by more than two orders of magnitude) than the dynamic length scale $\ell(t)$. That is, the pressure remains close to uniform in the (shrinking with time) collapse region. Still, the pressure steepening may signal a future breakdown of the uniform-pressure approximation at times extremely close to collapse, see Appendix B.

\section{Discussion}
\label{discussion}

We have found that,  in a circular container, the critical radius of the container for the clustering mode instability of a dilute nearly elastic granular gas is smaller than that for the shear mode instability. This is in contrast with rectangular geometry, where the shear mode bifurcates at a smaller system size than the clustering mode \cite{Goldhirsch,McNamara1,Brey,Brey2}. This fact enabled us to investigate a strongly nonlinear dynamics of a linearly unstable clustering mode in a 2D setting, without worrying about the shear mode.
We have presented analytic and numeric evidence for thermal collapse, unarrested by heat diffusion,
in a freely cooling dilute granular gas in 2D. Our results imply that Navier-Stokes hydrodynamics of dilute granular gases exhibits, in dimension higher than one, finite-time density blowups. This is in striking contrast to the ``wave breaking" singularities of an ideal flow (both in a ``classical" gas \cite{LL}, and in a granular gas \cite{Fouxon1,Fouxon2}) which are arrested by the viscosity and heat diffusion, bringing about \textit{smooth} shock-wave solutions.

We have derived the dynamic scaling laws of thermal collapse by assuming a locally uniform pressure profile. All the scaling laws, and the uniform-pressure scenario, have been corroborated in our hydrodynamic simulations.  An order-of-magnitude estimate, presented in Appendix B, suggests that the uniform-pressure scenario must break down sufficiently close to the singularity. We have not observed this breakdown in our hydrodynamic simulations: the uniform-pressure scenario agreed very well with the simulations until the latest simulation times we were able to reach. This may be related to some small numerical factors which improve the applicability of the uniform-pressure scenario beyond what our order-of-magnitude estimate predicts.   Future work should address this issue and find out whether the uniform-pressure scenario gives way to the zero-pressure scenario where a different type of singularity -- that of a free flow -- develops \cite{ELM,MP}.

It would be interesting to investigate the hydrodynamics of clustering in a  free cooling granular gas in 3D. One can speculate that the 2D singularities, analyzed in this work, are the \textit{generic} type of singularity in 3D. Indeed, a locally-2D pinching may be more common than a fully-3D collapse.

The dynamic scaling laws, that describe the density singularity, provide a sharp characterization of a strongly nonlinear stage of clustering instability. When the growing-in-time gas density peak reaches a fraction of the close-packing density $\sim \sigma^{-d}$, the dilute hydrodynamic equations break down. Our solutions describing thermal collapse represent, therefore, an \textit{intermediate} asymptotic of the true flow. Importantly, the duration of this intermediate asymptotic can be made arbitrary
long if the average density of the gas is made sufficiently small compared with the close-packing density.   Furthermore, the appearance of singularities in the dilute-hydrodynamic solutions signals the formation of closely-packed clusters of particles. The details of formation and dynamics of the closely-packed regions are obviously beyond a dilute-gas hydrodynamic description. A satisfactory description of granular gases at moderate densities (at a fraction of the close-packing density) is provided by a Navier-Stokes granular hydrodynamics with the
equation of state of Carnahan and Starling \cite{Carnahan} and
semi-empiric transport coefficients derived from
kinetic theory in the spirit of Enskog approach \cite{Jenkins}. A hydrodynamic description at even higher densities is even more challenging, although some particular flow regimes at very high densities have been successfully described by using empirical constitutive relations \cite{Grossman,Lubensky,MPB,Luding1} and, when necessary, by taking into account a multi-phase character of the flow \cite{KM,K}.

\section*{Acknowledgements}
We are very grateful to Clara Salue\~{n}a for extensive discussions which motivated a part of this
research.
We acknowledge useful discussions with Eran Sharon and Hillel Aharoni on the nature of shear flow in a circular geometry. We also thank Doron Kushnir for technical assistance.  This work was supported by the Israel Science Foundation.

\section*{Appendix A. Hydrodynamic simulations}
\renewcommand{\theequation}{A\arabic{equation}}
\setcounter{equation}{0}

We solved numerically a rescaled and transformed version of hydrodynamic equations Eqs.~(\ref{continuity})-(\ref{heat}). The rescaling reduces considerably the number of relevant parameters, whereas the transformation dramatically reduces the computation time.

The rescaling procedure is essentially the same we used at the beginning of section \ref{disk_section}: the coordinates $x,y$ are rescaled by the characteristic length scale $l_\kappa$ which appears in Eq.~(\ref{marginal_heat}), the time by the characteristic cooling time $t_c = 2/(\Lambda \rho_0^{1/2} p_0^{1/2})$, the gas velocity by $l_\kappa/t_c$, the density by the average density $\rho_0$, and the temperature by its typical initial value $T_0$. The rescaling does not change the continuity equation (\ref{continuity}). The momentum equation (\ref{NS}) becomes
\begin{equation}\label{NS_rescaled}
\epsilon_1\rho \left[ \frac{\partial \mathbf{v}}{\partial t} + (\mathbf{v}\cdot\mathbf{\nabla}) \mathbf{v} \right] = - \mathbf{\nabla} (\rho T) + \epsilon_2 \mathbf{\nabla} \cdot (\sqrt{T} \mathbf{\hat{\Sigma}}),
\end{equation}
where $\epsilon_1 = \kappa_0\Lambda/2$, and $\epsilon_2 = \nu_0\Lambda/2$. As we assume nearly-elastic collisions, see Eq.~(\ref{threeineq}), the dimensionless parameters $\epsilon_1 \sim \epsilon_2\sim 1-\alpha^2 \ll 1$.
The rescaled energy equation is
\begin{eqnarray}\label{heat_rescaled}
 \frac{\partial T}{\partial t} + (\mathbf{v}\cdot\mathbf{\nabla}) T  &=& -(\gamma - 1) T \mathbf{\nabla}\cdot \mathbf{v}  + \frac{1}{\rho} \mathbf{\nabla}\cdot ( \sqrt{T}\mathbf{\nabla}T) \nonumber \\  &-& 2 \rho T^{3/2} + \frac{\epsilon_2 (\gamma - 1) \sqrt{T}}{2 \rho} \mathbf{\hat{\Sigma}}^2\,.
\end{eqnarray}
In our simulation we assumed a 2D granular gas of hard disks, so $\gamma=2$.

The transformation we used is inspired by Haff's law. We introduce a transformed time, temperature and velocity:
\begin{eqnarray}
\tau = \ln(1+t),\nonumber\\
\tilde{T} =  (1+t)^2 T,\nonumber\\
\tilde{\mathbf{v}} = (1+t) \mathbf{v}\,,
\end{eqnarray}
whereas the (rescaled) density $\rho$ remains the same. The transformed equations become
\begin{equation}\label{continuity_t}
 \frac{\partial \rho}{\partial \tau} + \mathbf{\nabla}\cdot ( \rho \tilde{\mathbf{v}}) = 0
\end{equation}
\begin{eqnarray}\label{NS_t}
\epsilon_1\rho \left[ \frac{\partial \tilde{\mathbf{v}}}{\partial \tau}  + (\tilde{\mathbf{v}}\cdot\mathbf{\nabla}) \tilde{\mathbf{v}} \right] = &-& \mathbf{\nabla} (\rho T) + \epsilon_1 \rho \tilde{\mathbf{v}}\nonumber \\ &+&
\epsilon_2 \mathbf{\nabla} \cdot (\sqrt{T} \mathbf{\hat{\tilde{\Sigma}}}). \
\end{eqnarray}
\begin{eqnarray}\label{heat_t}
 \frac{\partial \tilde{T}}{\partial \tau} &+& (\tilde{\mathbf{v}}\cdot\mathbf{\nabla})  \tilde{T}  = -(\gamma - 1)  \tilde{T} \mathbf{\nabla}\cdot \tilde{\mathbf{v}}  + \frac{1}{\rho} \mathbf{\nabla}\cdot ( \sqrt{ \tilde{T}}\mathbf{\nabla} \tilde{T}) \nonumber \\  &-& 2 \rho  \tilde{T}^{3/2} + 2\tilde{T}+\frac{\epsilon_2 (\gamma - 1) \sqrt{ \tilde{T}}}{2 \rho}  \mathbf{\hat{\tilde{\Sigma}}}^2\,.
\end{eqnarray}
As one can see, two new source terms appear:  the
$\epsilon_1 \rho \tilde{\mathbf{v}}$ term in the transformed momentum equation, and the $2\tilde{T}$ term in the transformed energy equation. Neither of the source terms involves any derivatives, and so they could be easily implemented in our numerical codes. Importantly,
a time-dependent cooling state with a stationary density in the original formulation corresponds to
a \textit{true} steady state, for \textit{all} variables, in the transformed equations.

Two different hydrodynamic codes, Vulcan/1D (V1D) and Vulcan/2D (V2D), were used for a numerical solution of Eqs.~(\ref{continuity_t})-(\ref{heat_t}). The codes involve ALE (Arbitrary Lagrange-Euler) schemes on staggered grids and have both explicit and implicit solvers for hydrodynamics and
heat conduction \cite{Livne_codes}. V1D mainly works in the Lagrangian mode and therefore can efficiently handle 1D density singularities. V2D is naturally limited to Eulerian simulations. It can employ, however, a strongly non-uniform polar mesh. This feature of the code was very important when following the developing singularity in a circular container, see section \ref{hydrocircular}. The two codes were previously tested against other codes and used for a variety of astrophysical problems, as well as for granular hydrodynamic simulations \cite{LMS1,LMS2,Bromberg,ELM,Fouxon1,Fouxon2}.

\section*{Appendix B. Verifying assumptions \textit{a posteriori}}
\renewcommand{\theequation}{B\arabic{equation}}
\setcounter{equation}{0}
As we have seen in subsection \ref{Implications}, heat diffusion is irrelevant for thermal collapse in 2D or 3D. Let us check other assumptions that we made in order to derive the scaling relations for thermal collapse.

When using ideal equations, we neglected the viscous heating term in the energy equation (\ref{heat}) or (\ref{heatp}). The ratio of this term to the energy loss term can be estimated as
\begin{eqnarray}
\label{VH_vs_Cooling}
 \abs{ \frac{ \nu_{0} (\gamma - 1) \sqrt{p/\rho}\,\mathbf{\hat{\Sigma}}^2}{2\Lambda\rho^{1/2} p^{3/2}}} & \sim & \frac{\nu_0\Lambda}{2}\left(\frac{\rho}
 {\rho_0}\frac{p}{p_s}\right)^{-1}\left(\frac{v}{\ell} t_s \right)^2  \nonumber \\ &\sim & 1-\alpha^2 \ll 1\,.
\end{eqnarray}
Therefore, for nearly elastic collisions (that we assume throughout this work), the viscous heating remains small and, in the leading order, can be neglected.

The validity of hydrodynamics itself at the density blowup demands that the Knudsen number remains small, see Eq. (\ref{threeineq}). As we have seen in section \ref{scaling}, the smallest hydrodynamic scale varies with time, in 2D, as $\ell \sim (t_* - t)^{3/2}$. On the other hand, the mean free path of the gas varies as
$l_{free} \sim 1/\rho \sim (t_*-t)^2$. Therefore,
\begin{equation}
\frac{l_{free}}{\ell}\sim (t_*-t)^{1/2} \rightarrow 0\,,
\end{equation}
and the applicability of hydrodynamics improves near the singularity.

Finally, we will check whether the pressure profile remains locally uniform near the singularity. This amounts to checking whether the correction $p^{(1)}(\mathbf{r},t)$, which we neglected in the perturbative expansion (\ref{p_series}), is indeed much smaller than the term $p^{(0)}(t)$. To evaluate $p^{(1)}$ we return to Eq. (\ref{NS}) and plug in it the (supposedly leading-order) quantities found in section \ref{scaling}:
\begin{equation}
\pd{p^{(1)}}{r} = -\rho\left(\pd{v}{t}+v\pd{v}{r}\right)+\nu_0 \nabla \cdot \left(\sqrt{\frac{p^{(0)}}{\rho}} \,\mathbf{\hat{\Sigma}}\right)\,.
\end{equation}
Let us estimate the terms on the right hand side. The viscous term behaves, in the vicinity of the vertex of the container $r=0$, as
\begin{eqnarray}
\nu_0 &\nabla& \cdot(\sqrt{p^{(0)}/\rho}\, \mathbf{\hat{\Sigma}})\sim \frac{\nu_0}{\ell} \sqrt{p_s/\rho} \,\frac{v}{\ell}\nonumber \\ &\sim & \frac{\nu_0 \Lambda}{2\gamma}\frac{p_s}{\ell}\sim (1-\alpha^2)\frac{p_s}{\ell} \ll \frac{p_s}{\ell}\,.
\end{eqnarray}
As $\partial_r{p^{(1)}}\sim p^{(1)}/\ell$, the viscous term does not violate the strong inequality $p^{(1)} \ll p^{(0)}$.

Finally, let us estimate the inertial terms:
\begin{equation}
\rho\left(\pd{v}{t}+v\pd{v}{r}\right)\sim \frac{\rho v^2}{\ell}\sim  \kappa_0 \Lambda \frac{p_s}{\ell}\left(\frac{t_*-t}{t_s}\right)^{2\beta-4}.
\end{equation}
Therefore,  the contribution of the inertial terms to $p^{(1)}$ is such that
\begin{equation}
\frac{p^{(1)}}{p^{(0)}} \sim (1-\alpha^2)  \left(\frac{t_*-t}{t_s}\right)^{\frac{4}{d}-3}\,.
\end{equation}
One can see that, in the limit of nearly elastic collisions, $1-\alpha^2 \ll 1$, the ratio $p^{(1)}/p^{(0)}$ is small at early times. However, for $d=2$ or $3$ it grows without limit as $ t \rightarrow t_*$. This suggests that the uniform-pressure approximation breaks down sufficiently close to the time of singularity.

\end{document}